\documentstyle[epsfig]{elsart}
\begin{document}

\newcommand{\re}{\mathop{\mathrm{Re}}}
\newcommand{\im}{\mathop{\mathrm{Im}}}
\newcommand{\D}{\mathop{\mathrm{d}}}
\newcommand{\I}{\mathop{\mathrm{i}}}
\newcommand{\E}{\mathop{\mathrm{e}}}

\noindent {\Large DESY 02-127}

\noindent {\Large September 2002}

\small

\begin{frontmatter}
 
\journal{Journal of Synchrotron Radiation}

\date{}

\title{
Femtosecond Resolution Experiments at Third-Generation Light Sources:  
a Concept Based on the Statistical Properties of Synchrotron 
Radiation} 

\author[DESY]{E.L.~Saldin}, 
\author[DESY]{E.A.~Schneidmiller},
\author[Dubna]{M.V.~Yurkov}

\address[DESY]{Deutsches Elektronen-Synchrotron (DESY), 
Notkestrasse 85, D-22607 Hamburg, Germany}

\address[Dubna]{Joint Institute for Nuclear Research, Dubna, 
141980 Moscow Region, Russia}

\begin{abstract}

The paper describes a new concept of visible pump/X-ray probe/slow 
detector experiments that could be performed at third-generation 
synchrotron light sources. We propose a technique that would allow time 
resolution up to femtosecond capabilities to be recovered from a long 
(100 ps) X-ray probe pulse. The visible pump pulse must be as short as 
the desired time resolution. The principle of operation of the proposed 
pump-probe scheme is essentially based on the statistical properties of 
the synchrotron radiation. These properties are well known in 
statistical optics as properties of completely chaotic polarized light.  
Our technique utilizes the fact that, for any synchrotron light beam 
there exist some characteristic time (coherence time), which determines 
the time-scale of the random fluctuations. The typical coherence time 
of soft X-ray synchrotron light at the exit of monochromator is in 
the femtosecond range. An excited state is prepared with a pump pulse 
and then projected with a probe pulse onto a final ion state. The 
first statistical quantity of interest is the variance of the number of 
photoelectrons detected during synchrotron radiation pulse.  The 
statistics of concern are defined over an ensemble of synchrotron 
radiation pulses.  From a set of variances measured as a function of 
coherence time (inversely proportional to monochromator bandwidth) it 
is possible to reconstruct the femtosecond dynamical process. 

\end{abstract}

\end{frontmatter}

\clearpage

\setcounter{page}{1}

\section{Introduction}

Time-resolved experiments are used to monitor time-dependent phenomena.
The study of dynamics in physical systems often requires time 
resolution beyond the picosecond capabilities presently available with 
synchrotron radiation. Femtosecond ($10^{-15}$s) capabilities have been 
available for many years at visible wavelengths. There exists a wide 
interest in the extension of femtosecond techniques to the soft X-ray, 
and X-ray regions of spectrum. 

\begin{figure}[b]
\begin{center}
\epsfig{file=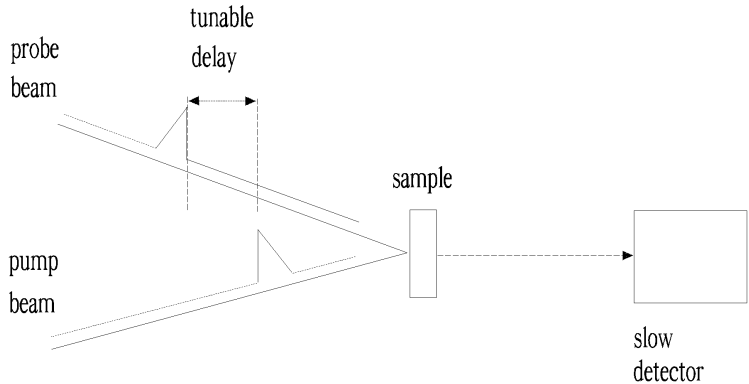,width=0.75\textwidth}
\end{center}
\caption{
Typical scheme of a "short pump-short probe" experiment} 
\label{fig:pps1} 
\end{figure}

The time resolution of pump-probe experiment is, obviously, determined 
by the duration of the pump as well as the resolution of the probing. 
The pump pulse must always be as short as the desired time resolution. 
In typical scheme of a pump-probe experiment (see Fig. \ref{fig:pps1}) 
the short probe pulse follows the pump pulse at some specified delay. 
The signal recorded by the slow detector then reflects the state of the 
sample during the brief probing. The experiment must be repeated many 
times with different delays in order to reconstruct the femtosecond 
dynamical process \cite{f1,f2}. 

\begin{figure}[tb]
\begin{center}
\epsfig{file=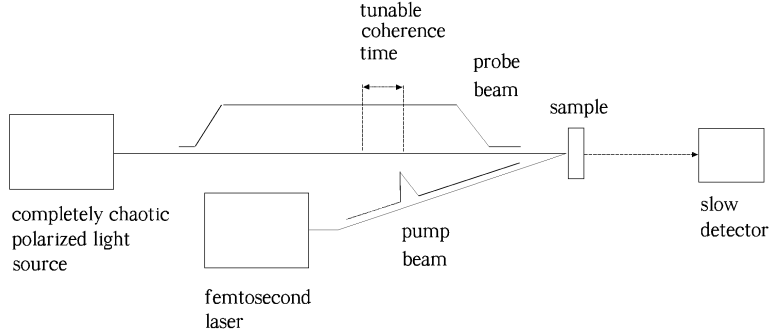,width=0.75\textwidth}
\end{center}
\caption{
The new scheme of "pump-probe" experiment. The principle of operation 
of the proposed scheme is based essentially on the statistical 
properties of synchrotron radiation, which possesses all the features 
corresponding completely chaotic polarized light} 
\label{fig:pps2} 
\end{figure}

\begin{figure}[tb]
\begin{center}
\hspace*{0.01\textwidth}
\epsfig{file=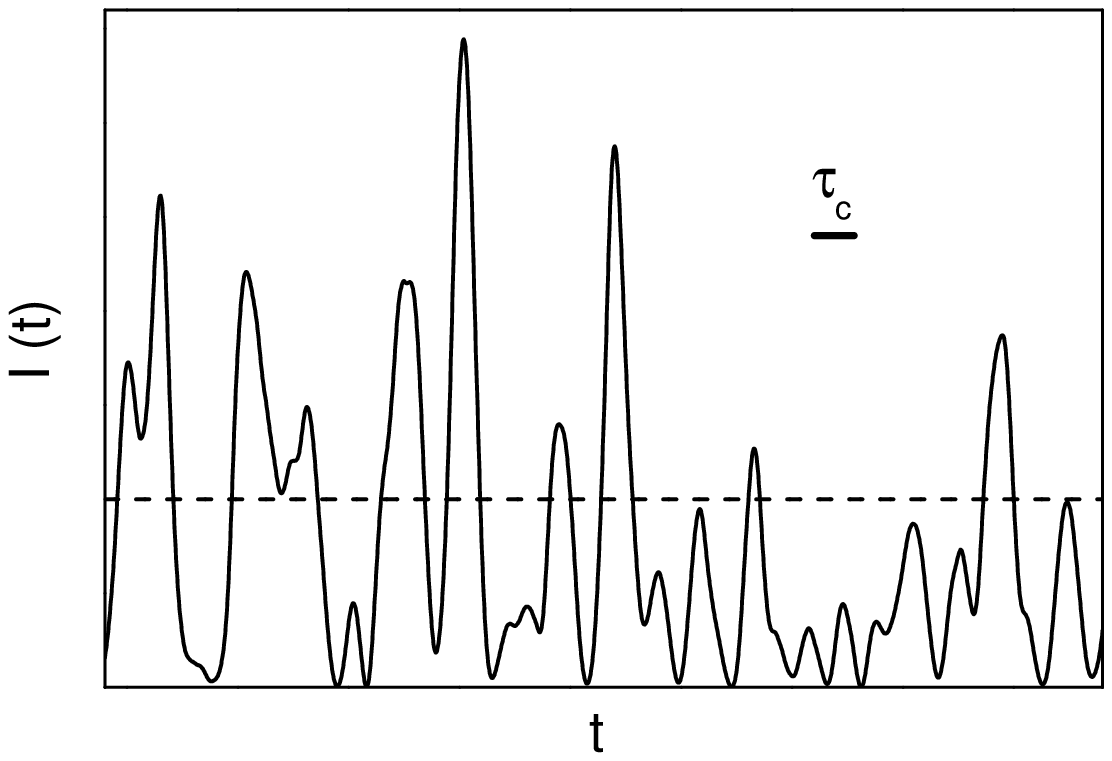,width=0.765\textwidth}

\vspace*{-10mm}

\epsfig{file=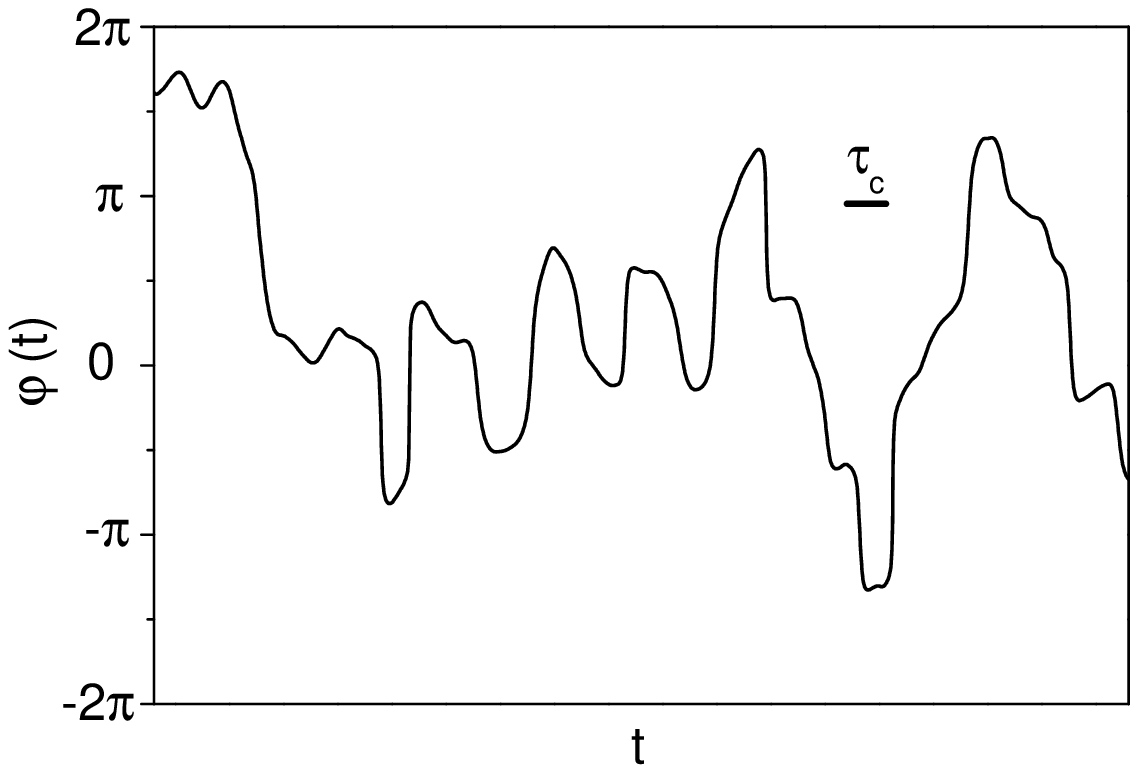,width=0.8\textwidth}
\end{center}
\caption{
Time-dependence of the intensity (upper plot) and phase (lower plot)
for synchrotron light, obtained from a 
computer simulation.The dashed line shows the mean value of the 
intensity averaged over an ensemble of radiation pulses 
} 
\label{fig:int} 
\end{figure}

The obvious temporal limitation of the typical visible pump/X-ray probe 
technique is the duration of the X-ray probe. Here we will 
concentrate on the performance obtained with 
third-generation storage rings.  At these sources, the X-ray pulse 
duration is about 100 ps. The new principle of pump-probe techniques 
described below offers a way around this difficulty. 
We propose visible pump / 
X-ray probe technique that would allow time resolution up to 
femtosecond capabilities to be recovered from a long X-ray probe pulses 
(see Fig. \ref{fig:pps2}).  The principle of operation of the proposed 
pump-probe scheme is essentially based on the statistical properties of 
the synchrotron radiation. Synchrotron radiation is a stochastic object 
and at a given time it is impossible to predict the amount of energy 
which flows to a sample. For any synchrotron light beam there exist 
some characteristic time, which determines the time-scale of the random 
fluctuations. This characteristic time is called, in general, the 
coherence time $\tau_{\mathrm{c}}$ of the synchrotron light beam. Its 
magnitude is of the order of the inverse of the frequency spread of the 
beam.  In all the theory that follows, attention is restricted to 
synchrotron light beams whose frequency spreads are small compared with 
the mean frequency, that is, where $\omega_{0}\tau_{\mathrm{c}}$ is 
very much larger than unity.  Figure \ref{fig:int} 
illustrates the type of fluctuations that occur in 
the synchrotron radiation beam intensity and phase. The figures have 
been constructed by a computer simulation of a synchrotron light source 
in which the summation is carried out explicitly for the real number of 
electrons $(N \simeq 10^{10})$ in the electron bunch.  It is seen that 
substantial changes in intensity (and phase) occur over a time span 
$\tau_{\mathrm{c}}$, but these quantities are reasonably constant 
over time intervals $\Delta t \ll \tau_{\mathrm{c}}$.     

\noindent Our technique utilizes the 
fact that, over the long probe pulse, time correlations exist   
in the  probe pulse intensity. 
It was shown that the femtosecond timescale associated with intensity 
fluctuations in synchrotron light pulses make them well suited for 
time-resolved studies. The first statistical quantity of 
interest is the variance of the detector counts.   
The statistics of concern are defined over an ensemble of synchrotron 
radiation pulses. From a set of variances measured 
as a function of correlation time it is possible to reconstruct the 
femtosecond dynamical process. 

\section{The principle of pump-probe techniques based on the 
statistical properties of synchrotron light}

The shot noise in the electron beam causes fluctuations of the beam 
density which are random in time and space. As a result, the 
radiation, produced by such a beam, has random amplitudes and phases in 
time and space. These kinds of radiation fields can be described in 
terms of statistical optics, branch of optics that has been developed 
intensively during the last few decades and there exists a lot of 
experience and a theoretical basis for the description of fluctuating 
electromagnetic fields \cite{l,g}.

We will begin in this section by dealing with some general notions of 
statistical optics. Some of the statements will be quite precise, other only 
partially precise. 
In the framework of statistical optics the 
radiation field is characterized by notions such as time and space 
coherence. Let us illustrate these notions. To be specific, we consider 
the radiation pulse of nearly monochromatic radiation having a duration 
and bandwidth equal to $T$ and $\Delta\omega$, respectively. If $T \gg 
1/\Delta\omega$, the radiation is partially coherent in time. The time 
coherence is of about $\tau_{\mathrm{c}} \simeq 1/\Delta\omega$. 
The physical sense of this notion is as follows. Let us separate the 
radiation pulse by a splitter. These two pulses pass different path 
lengths and are then combined together. If the difference between the 
path lengths is less than $c\tau_{\mathrm{c}}$, we see the 
interference pattern at the end for a shot-to-shot averaging (averaging 
over a large number of radiation pulses). Using simple physical 
language, we can say that the radiation field is correlated within the 
time of coherence.         

The notion of space coherence can be explained in the same way. Let us 
direct the radiation onto a screen with two pinholes and look at the 
interference pattern in the far diffraction zone when changing the 
distance between the pinholes. When the pinholes are located close to 
each other, we see a clear diffraction pattern. As the distance between 
the holes increases to some value, $D > \Delta r_{\mathrm{c}}$, we 
come to the situation when we do not see an interference pattern after 
averaging over the ensemble of pulses. Qualitatively, the value of 
$\pi(\Delta r_{\mathrm{c}})^{2}$ is referred as the area of coherence 
of the radiation pulse. The coherence volume is defined as the product 
of the coherence area, $\pi(\Delta r_{\mathrm{c}})^{2}$, with the 
coherence length, $c\tau_{\mathrm{c}}$. If one uses the notions of 
quantum mechanics, this volume corresponds to one cell in the phase 
space of the photons.  The number of photons in the coherence volume is 
also referred to as the degeneracy parameter, $\delta_{\mathrm{W}}$. 
Physically this parameter means the average number of photons which can 
interfere, or the number of photons in one quantum state (one mode). It 
should be noted that third-generation synchrotron light sources can 
produce soft X-ray radiation with very high degeneracy parameter and in 
such cases we can state that the classical approach is adequate for a 
description of the statistical properties of the synchrotron radiation.

Electron beam is composed of large number of electrons, thus 
fluctuations always exist 
in the electron beam density due to the effect of shot noise. When the 
electron beam enters the undulator, the presence of the beam modulation 
at frequencies close to the resonance frequency initiates the process 
of radiation. The shot noise in the electron beam is a Gaussian random 
process.  The monochromator can be considered as a linear filter which 
does not change the statistics of the signal. As a result, we can 
define general statistical properties of the radiation after 
monochromator without any calculations. For instance, the 
real and imaginary parts of the slowly varying complex amplitudes of 
the electric field of the electromagnetic wave have a Gaussian 
distribution, the instantaneous radiation power fluctuates in 
accordance with the negative exponential distribution, and the energy 
in the radiation pulse fluctuates in accordance with the gamma 
distribution. We can also state that the spectral density of the 
radiation energy and the first-order time correlation function should 
form a Fourier transform pair (this is so-called Wiener Khintchine 
theorem). Also, the higher-order correlation functions (time and 
spectral) should be expressed in terms of the first-order correlation 
functions. These properties are well known in statistical optics as 
properties of completely chaotic polarized radiation \cite{g}.
This demonstrate one of the most beautiful things about statistical 
optics - how much can be deduced from so little.  This 
brings up an interesting question: Why is it that the noise in the 
electron bunch is a Gaussian random process? We will give an 
explanation in section 3.

\begin{figure}[tb]
\begin{center}
\epsfig{file=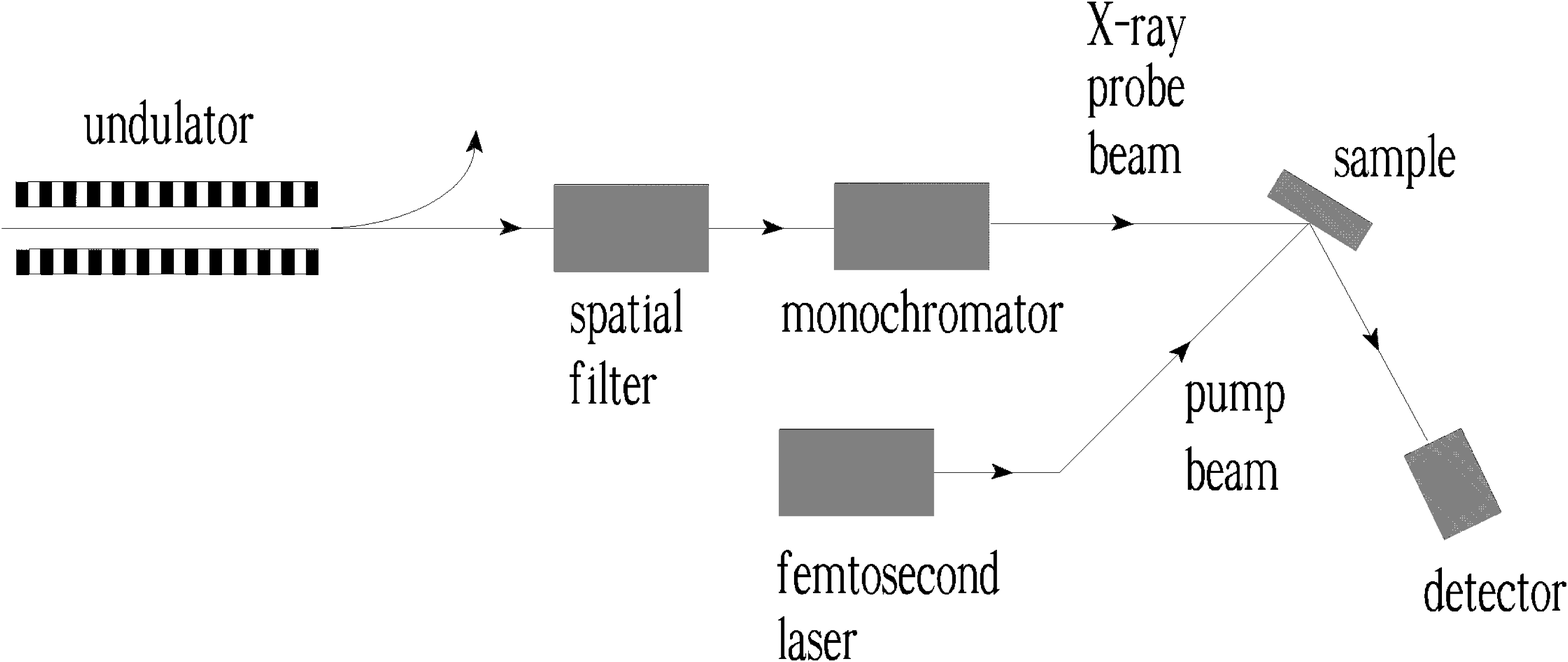,width=0.75\textwidth}
\end{center}
\caption{
Proposed pump-probe experimental setup  
} 
\label{fig:pps5} 
\end{figure}

The arrangement of components for a long probe pulse experiment is 
shown in Fig. \ref{fig:pps5}. In order to use the radiation from 
synchrotron light source it first has to be spatially filtered and 
monochromatized.  We assume that synchrotron radiation reaching the 
sample surface has full transverse coherence. With this assumption, 
attention can be concentrated completely on temporal coherence effects.  
The coherence time of synchrotron light is 
inversely proportional to the monochromator bandwidth. So the narrower 
we make the monochromator exit slit, the wider the coherence time gets.  
Any measurement of photodetector counts will be 
accompanied by certain unavoidable fluctuations and we wish to 
determine the statistical distribution of the number of photocounts 
observed in any radiation pulse.

\begin{figure}[tb]
\begin{center}
\epsfig{file=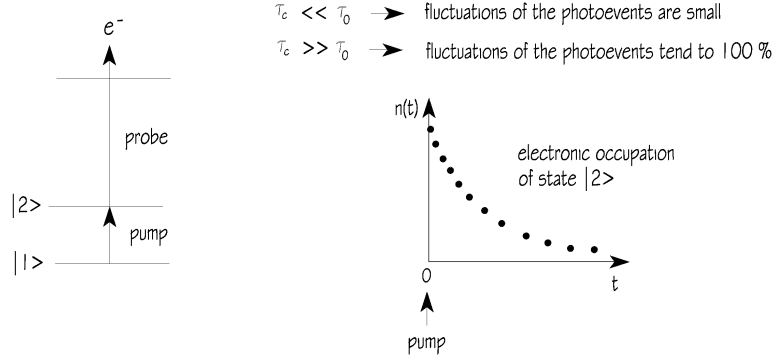,width=0.75\textwidth}
\end{center}
\caption{
Schematic ionization scheme for pump-probe ionization in the visible 
pump/X-ray probe transition. A typical 
experiment would be to couple two electronic state $|1\rangle$ and 
$|2\rangle$ resonantly by a femtosecond pump-field. The dynamics is 
then probed by ionization and detection of the fluctuations of 
photoelectron counts per pulse as a function of the coherence time of 
the probe synchrotron radiation pulse} 
\label{fig:pps8} 
\end{figure}

Now we wish to consider one example which shows the relationship 
between the fluctuations of photocounts and the coherence time in a 
circumstances that easy to understand. We will not discuss applications 
of the proposed pump-probe scheme.  There are, in fact, many 
applications that could be discussed.  We have chosen for emphasis here 
experiments aimed at measuring the lifetime of electronic excitation in 
condensed matter.  For our present purposes we would like to imagine a 
somewhat idealized experiment. 
Several assumptions are made about the process of 
internal conversion of the electronic excitation.  First, the time 
dependence of the electronic occupation is assumed to be known a priory 
-- this is the exponential dependence $n(t) = n_{0}\exp(-t/\tau_{0})$.  
In practice, this is a good assumption.   The purpose of the 
measurement is presumably to determine the lifetime $\tau_{0}$. The 
typical lifetime of the electronic excitation is in the femtosecond 
range.  This type of time-resolved experiments may be understood as 
preparing an excited state with a pump pulse and then projecting it 
with a probe pulse onto a final state.  The final state should be well 
characterized.  Photoionization detection has several advantages. The 
ion state may be well characterized by independent methods such as high 
resolution photoelectron spectroscopy.  Schematic ionization scheme for 
pump-probe ionization is illustrated in Fig. \ref{fig:pps8}. A typical 
experiment would be to couple two electronic state $\mid 1\rangle$ and 
$\mid 2\rangle$ resonantly by a femtosecond pump-field at a frequency 
$\omega$ in the visible range.  The dynamics is then probed by 
ionization and detection of the photoelectrons as a function of the 
coherence time of the probe synchrotron radiation pulse. A 
time-resolved experiment inherently begins with the initiation of the 
process under study, at some more or less accurately defined instant in 
time. We assume that the pump pulse can be approximated by a 
$\delta$-function. With that simplifying assumption, the number of 
photoelectrons $K$ detected during a synchrotron radiation pulse is 
directly proportional to the integrated value of the physical parameter 
(occupation number) to be measured: 

\begin{displaymath}
K \propto \int\limits^{\infty}_{0}I(t)n_{0}\exp(-t/\tau_{0})\D t \ ,
\end{displaymath}

\noindent where $I(t)$ is the instantaneous intensity of synchrotron 
radiation. At time $t = 0$ the pump pulse perturbs the sample. Here
we assume that the pump-probe process under 
consideration can be separated from what might be called parasitic 
processes.  In practice, not every count will be detected, and some 
false counts will register. We are not considering such details because 
our interest is in the fundamental aspects of the problem. 

A quantity of considerable physical interest is the variance of the 
photocount distribution

\begin{displaymath}
\sigma^{2}_{\mathrm{K}} = \frac{\langle K^2\rangle - \langle 
K\rangle^{2}}{\langle K\rangle^{2}} \ .   
\end{displaymath}

\noindent Determination of the variance $\sigma_{\mathrm{K}}^{2}$ 
as a function of the monochromator bandwidth 
$\Delta\omega_{\mathrm{m}}$ gives us information on the lifetime 
$\tau_{0}$.  Why?  First, for a lifetime $\tau_{0}$ that is much 
shorter than the coherence time $\tau_{\mathrm{c}}$ of the probe pulse, 
the number of photoelectrons is to an excellent approximation, simply 
the product of the instantaneous intensity and the lifetime $\tau_{0}$,

\begin{displaymath}
K \propto \int\limits^{\infty}_{0} I(t)n_{0}\exp(-t/\tau_{0})\D t 
\simeq I(0)n_{0}\tau_{0}  \qquad {\mathrm{for}} \quad 
\Delta\omega_{\mathrm{m}}\tau_{0} \ll 1 \ . 
\end{displaymath}

\noindent Within the scaling factor, therefore, the probability density 
function of $K$ is approximately the same as the density function of 
the instantaneous intensity. Using the well-known results obtained in 
the framework of statistical optics, we can state that when the 
lifetime is much shorter than the inverse of the monochromator 
bandwidth, the variance tends to unity:

\begin{displaymath}
\sigma^{2}_{\mathrm{K}} \simeq 1 
\qquad {\mathrm{for}} \quad 
\Delta\omega_{\mathrm{m}}\tau_{0} \ll 1 \ . 
\end{displaymath}

\begin{figure}[tb]
\begin{center}
\epsfig{file=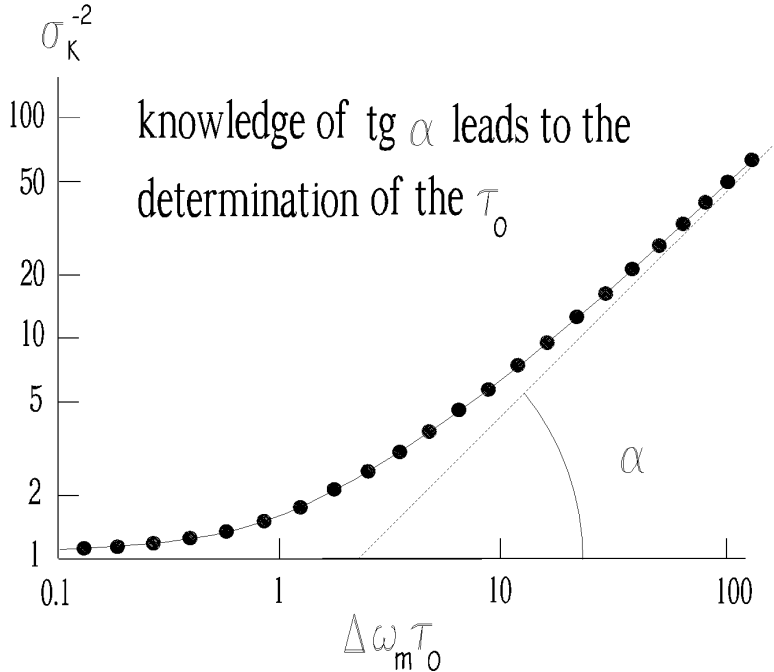,width=0.75\textwidth}
\end{center}
\caption{
Typical behavior of the variance associated with the measurement of 
the number of photoelectrons per pulse as a function of lifetime and 
monochromator bandwidth} 
\label{fig:pps6} 
\end{figure}

At the opposite extreme, with a lifetime much 
longer than the coherence time, the fact that many independent 
fluctuations of the instantaneous intensity occur within the interval 
$\tau_{0}$ implies, according to the theory of completely chaotic 
light, that the variance is inversely 
proportional to the monochromator bandwidth. So, we obtain the rule 
that the product of lifetime and monochromator bandwidth is of the 
order of 

\begin{displaymath}
\Delta\omega_{\mathrm{m}}\tau_{0} \simeq 
\sigma^{-2}_{\mathrm{K}} 
\qquad {\mathrm{for}} \quad     
\Delta\omega_{\mathrm{m}}\tau_{0} \gg 1 \ . 
\end{displaymath}

\noindent This relation is illustrated in Fig. \ref{fig:pps6}, which 
shows the parameter $\sigma^{-2}_{\mathrm{K}}$ plotted against 
$\Delta\omega_{\mathrm{m}}\tau_{0}$. 
The exact results can in fact be found for certain monochromator line 
shapes using the formalism introduced in Section 2. The dependence of 
$\sigma^{-2}_{\mathrm{K}}$ on the exact shape of the signal and of the 
monochromator profile is rather weak and can be ignored outside the 
range $0.1 < \Delta\omega_{\mathrm{m}}\tau_{0} < 10$.

It is important to note that the photocount fluctuations discussed 
above originate from the shot noise in the electron beam. Nevertheless, 
the problem of photocount fluctuations is more complicated and there is 
another (parasitic) fluctuation effect that can be important for 
the proposed pump-probe experiments. The problem is the time jitter of 
electron bunch and pump pulse. Any measurement of photocounts will be 
accompanied by these unavoidable fluctuations.  To solve this problem, 
femtosecond pump pulses from the laser system should be synchronized to 
the master clock of the storage ring rf system with phase-locking 
technique. At present the best achievable synchronization between the 
optical laser and the synchrotron light source is of the order of a few 
picoseconds. On the other hand the electron bunch length is about of 
100 ps. Assuming that the electron bunch envelope is Gaussian 
distribution (in practice, this is a good assumption) we can find that 
the parasitic photocount variance reaches a value of about a per cent 
only.  Thus we find that pump pulse jitter should not be a serious 
limitation in our case.

The next problem is to define the sensitivity of the proposed 
pump-probe method.  A general study of 
the noise fluctuations associated with the photoevents would 
be nontrivial. The difficulty arises in simultaneously including the 
effects of both the classically induced fluctuations  and 
Poisson noise fluctuations associated with the basic interaction of 
light and matter.  The ratio of the classical variance to the "photon 
noise" variance is named as a photocount degeneracy parameter 
$\delta_{\mathrm{c}}$ (see section 4). Physically speaking, the count 
degeneracy parameter can be interpreted as the average number of counts 
that occur in a single coherence interval of the incident (transversely 
coherent) radiation.  It can also be described as the average number of 
counts per "degree of freedom" or per "mode" of the incident wave. When 
$\delta_{\mathrm{c}} \ll 1$, it is highly probable that there will be 
no more than one count per coherence interval of the wave, with the 
result that quantum noise predominates over classically induced noise. 
On the other hand, when $\delta_{\mathrm{c}} \gg 1$, there are many 
photoevents presented in each coherence interval of the wave. The 
result is a "bunching" of the photoevents by the classical intensity 
fluctuations, and an increase of the variance of the counts to the 
point where the classically induced fluctuations are far stronger than 
the quantum noise variations.      

Let us consider the case when the degeneracy parameter is 
much larger than unity, $\delta_{\mathrm{c}} \gg 1$. The fluctuations 
of the photoevents are defined mainly by the classical noise in this 
case and the signal-to-noise ratio for the proposed device can be 
written in the form

\begin{equation}
\frac{S}{N} = \frac{\sqrt{N_{\mathrm{shot}}}\langle(K - \langle 
K\rangle)^{2}\rangle}{\sqrt{\langle (K - \langle K\rangle)^{4}\rangle -
\langle (K - \langle K\rangle)^{2}\rangle^{2}}} \qquad 
{\mathrm{for}}\qquad \delta_{\mathrm{c}} \gg 1 \ , 
\label{eq:sn}
\end{equation}

\noindent where $N_{\mathrm{shot}}$ is the number of independent 
measurements averaged in the accumulator (total number of shots). 
The determination of the signal-to-noise ratio is a nontrivial problem. 
Fortunately, in the particular case, namely, 
$\tau_{\mathrm{c}} \gg \tau_{0}$, a much simplified analysis is 
sufficient. In this case the expression  (\ref{eq:sn}) can be reduced to

\begin{displaymath}
S/N \simeq \sqrt{N_{\mathrm{shot}}/8} \qquad 
{\mathrm{for}} \qquad \delta_{\mathrm{c}} \gg 1 \ , \quad    
\Delta\omega_{\mathrm{m}}\tau_{0} \ll 1 \ .
\end{displaymath}

\noindent The signal-to-noise ratio depends on the degeneracy parameter 
when $\delta_{\mathrm{c}}$ is much less than unity. In this case 
one can derive that

\begin{displaymath}
S/N \simeq 
\delta^{3/2}_{\mathrm{c}}\sqrt{N_{\mathrm{shot}}} \qquad
{\mathrm{for}} \qquad \delta_{\mathrm{c}} \ll 1 \ , \quad    
\Delta\omega_{\mathrm{m}}\tau_{0} \ll 1 \ .
\end{displaymath}

The significance of the result we have obtained cannot be fully 
appreciated until we determine typical values of the degeneracy 
parameter that can be expected in practice. 
Let us consider the 
case of radiation with a sufficiently long coherence time that 
$\tau_{\mathrm{c}} \gg \tau_{0}$. Under such condition, 
the photocount degeneracy parameter can be estimated simply as (see 
section 5)

\begin{equation}
\delta_{\mathrm{c}} \simeq \eta(\Delta\omega_{\mathrm{m}}\tau_{0}) 
\left(R_{\mathrm{m}}\frac{\lambda^{3}B_{\mathrm{peak}}}{4c}\right)
\qquad {\mathrm{for}} \qquad 
\Delta\omega_{\mathrm{m}}\tau_{0} \ll 1 \ ,   
\label{eq:sn1}
\end{equation}

\noindent where $\eta$ is the quantum efficiency of photoelectron 
production, $R_{\mathrm{m}}$ is the monochromator throughput, $\lambda$ 
is the radiation wavelength, 
$B_{\mathrm{peak}}$ is the peak spectral brightness. 
Our physical interpretation of the result (\ref{eq:sn1}) is as follows. 
In order to use the radiation from the undulator it first has to be 
spatially filtered and monochromatized (see Fig. \ref{fig:pps5}). We 
interpret the factor $R_{\mathrm{m}}\lambda^{3}B_{\mathrm{peak}}/(4c)$ 
as representing the number of photons per coherence interval at the 
monochromator exit, whereas the factor 
$\Delta\omega_{\mathrm{m}}\tau_{0}$ represents a ratio of 
integration time $\tau_{0}$ to coherence 
interval $\tau_{\mathrm{c}} \simeq 1/\Delta\omega_{\mathrm{m}}$. 
Let us present a specific numerical example for the case of a 
third-generation synchrotron light source.  The spectral brightness 
delivered by an undulators at a wavelength of $\lambda \simeq 10$ nm 
is about $B_{\mathrm{peak}} \simeq 10^{21}$ 
photons/s/0.1\%BW/${\mathrm{mm}}^{2}/{\mathrm{mrad}}^{2}$ 
and the degeneracy parameter is about $\delta_{\mathrm{c}} \simeq 
10^{3}\eta R_{\mathrm{m}}$ for the case when $\tau_{\mathrm{c}} \simeq 
\tau _{0}$.

As we mentioned above, the dependence of 
$\sigma^{-2}_{\mathrm{K}}(\tau_{\mathrm{c}})$ on the exact 
shape of the signal is rather weak. As a result, the  
behaviour of the variance as a function of synchrotron radiation 
coherence time provides only very little information on the  
sample dynamics. Nevertheless, the described pump-probe technique can 
be used to quote a "characteristic time" of the sample dynamics. The 
typical procedure is to assume a signal shape (generally exponential 
shape), and to "determine" the relaxation time $\tau_{0}$ from the 
known ratio between the $\tau_{0}/\tau_{\mathrm{c}}$ and 
$\sigma_{\mathrm{K}}^{2}$.  

A more complex instrument can be used to characterize the shape of the 
signal function. The pump-probe technique based on a correlation 
principle we are describe in section 7 is to use two spatially 
separated samples and record correlation of the count fluctuations for 
each value of the delay $\tau$ between two pump laser pulses. Averaged 
(shot to shot) count product contains information about the shape 
of the signal function. Another advantage of the correlator measurement 
is the possibility to remove the monochromator between spatial filter 
and sample and thus to increase the count degeneracy parameter.

\section{Statistical properties of synchrotron radiation}

The present section considers the statistical properties of the 
intensity fluctuations in synchrotron light. In order to give the 
subject a semblance of continuity, it will be desirable to introduce 
considerable matter which will be found in any of standard works on 
statistical optics. Much of the credit for stimulating modern 
development in statistical analysis of synchrotron radiation is due to 
Professor's Goodman book \cite{g}, which laid the groundwork for 
the principles of pump-probe techniques based on statistical properties 
of synchrotron light. Quite basic similarities between the methods 
described here and the methods introduced into X-ray free electron 
laser physics by authors in \cite{k} should be recognized. We here 
consider the classical theory of the fluctuation experiment, reserving 
the quantum discussion for section 4, where the theory of intensity 
fluctuations is reconsidered in terms of the quantized radiation field.

\subsection{Shot noise in the electron beam}

In this subsection we study the statistical properties of the 
shot noise in the electron beam. It should be noted that the process 
under study is nonstationary with finite pulse duration, so in what 
follows the averaging symbol $\langle\cdots\rangle$ means the ensemble 
average over bunches.

Let us consider the microscopic picture of the electron beam current at 
the entrance of the undulator. The electron beam current is made up of 
moving electrons randomly arriving at the entrance to the undulator:

\begin{displaymath}
J(t) = (-e)\sum^{N}_{k=1}\delta(t-t_{k}) \ ,
\end{displaymath}

\noindent where $\delta(\cdots)$ is the delta function, $(-e)$ is the 
charge of the electron, $N$ is the number of electrons in a bunch, and 
$t_{k}$ is the random arrival time of the electron at the undulator 
entrance. The electron bunch profile is described by the profile 
function $F(t)$. The beam current averaged over an ensemble of bunches 
can be written in the form:

\begin{displaymath}
\langle J(t)\rangle = (-e)NF(t) \ .      
\end{displaymath}

\noindent For instance for an electron beam with Gaussian distribution 
of the current along the beam, the profile function is:

\begin{displaymath}
F(t) = \frac{1}{\sqrt{2\pi}
\sigma_{\mathrm{T}}}\exp\left(-\frac{t^{2}}
{2\sigma^{2}_{\mathrm{T}}}\right) \ . 
\end{displaymath}

\noindent The probability of arrival of an electron during the time 
interval $(t,t+\D t)$ is equal to $F(t)\D t$.

The electron beam current, $J(t)$, and its Fourier transform, 
$\bar{J}(\omega)$, are connected by

\begin{equation}
\bar{J}(\omega) = \int\limits^{\infty}_{-\infty}e^{\I\omega t}J(t)\D t 
= (-e)\sum^{N}_{k=1}e^{\I\omega t_{k}} \ , 
\label{eq:s1a}
\end{equation}

\begin{figure}[tb]
\begin{center}
\epsfig{file=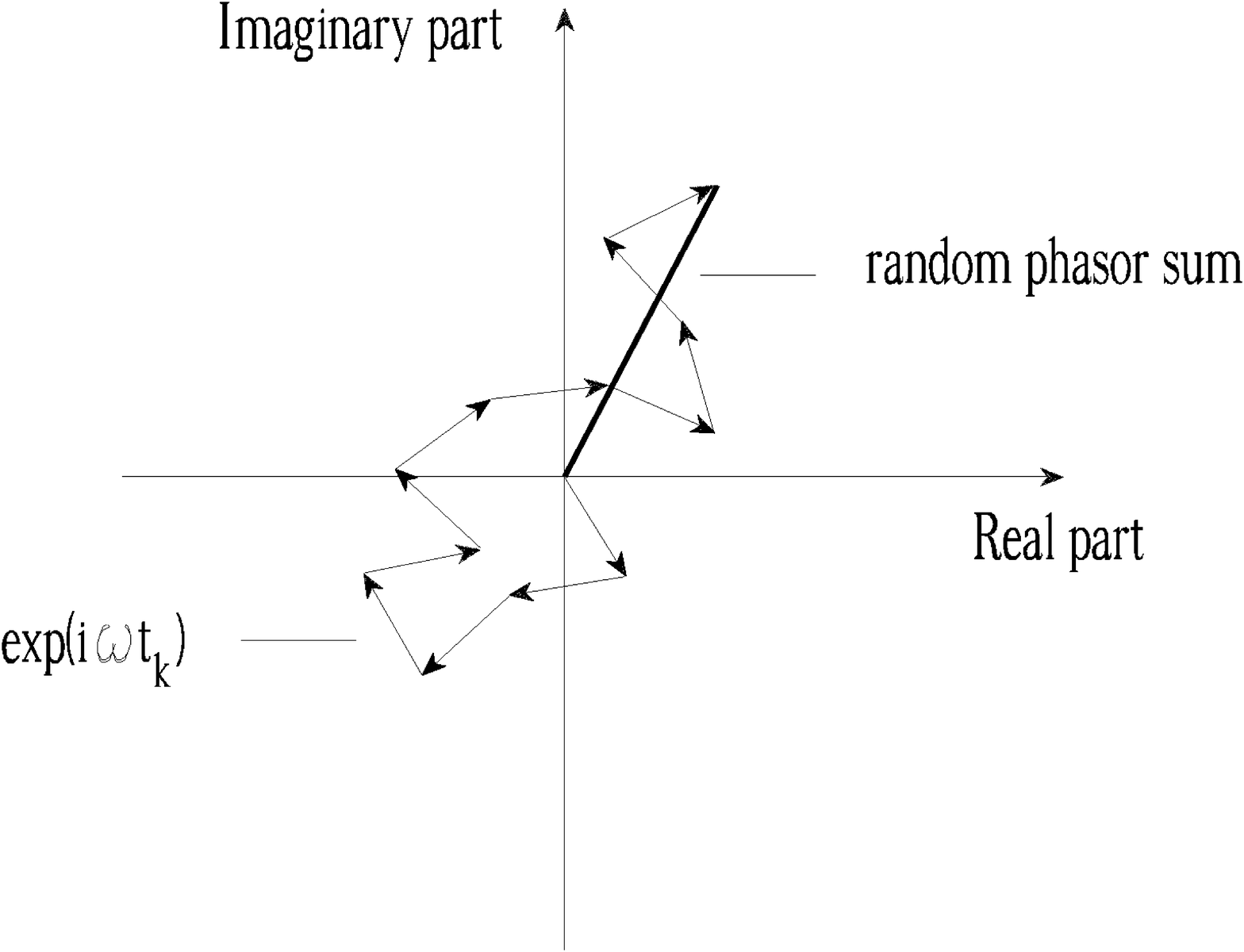,width=0.75\textwidth}
\end{center}
\caption{
Diagram to show the amplitude and phase of the resultant vector (total 
complex amplitude) formed by a large number of unit vectors 
(elementary complex-valued contributions $\exp(\I\omega t_{k})$), each 
of which has a randomly chosen phase angle} 
\label{fig:pps7} 
\end{figure}

\begin{displaymath}
J(t) = 
\frac{1}{2\pi}\int\limits^{\infty}_{-\infty}\bar{J}(\omega)e^{-\I\omega 
t}\D\omega = (-e)\sum^{N}_{k=1}\delta(t - t_{k}) \ . 
\end{displaymath}

\noindent It follows from these expressions that the Fourier transform 
of the input current, $\bar{J}(\omega)$, is the sum of a large number 
of complex phasors with random phases $\phi_{k} = \omega t_{k}$. When 
the electron pulse duration, $\sigma_{\mathrm{T}}$, is long, 
$\omega\sigma_{\mathrm{T}} \gg 1$, the phases $\phi_{k}$ can be 
regarded as uniformly distributed on the interval $(0,2\pi)$. 
The formal summation of the phasors is illustrated in Fig. 
\ref{fig:pps7}. The calculation is an example of the "random walk" 
problem, well known in the theory of stochastic process. 
In this 
case we can use the central limit theorem and conclude that the real 
part and the imaginary part of $\bar{J}(\omega)$ are distributed in 
accordance with the Gaussian law \cite{g}. The probability density 
distribution of $\mid\bar{J}(\omega)\mid^{2}$ is given by the negative 
exponential distribution:

\begin{equation}
p(\mid\bar{J}(\omega)\mid^{2}) = 
\frac{1}{\langle\mid\bar{J}(\omega)\mid^{2}\rangle} 
\exp\left(-\frac{\mid\bar{J}(\omega)\mid^{2}}{
\langle\mid\bar{J}(\omega)\mid^{2}\rangle}\right) \ . 
\label{eq:s1}
\end{equation}

\subsection{First-order spectral correlation}

Let us calculate the first-order correlation of the complex Fourier 
harmonics $\bar{J}(\omega)$ and $\bar{J}(\omega^{\prime})$:

\begin{displaymath}
\langle \bar{J}(\omega)\bar{J}^{*}(\omega^{\prime})\rangle = 
e^{2}\langle \sum^{N}_{k=1}\sum^{N}_{n=1}\exp(\I\omega t_{k} - 
\I\omega^{\prime} t_{n})\rangle \ .  
\end{displaymath}

\noindent Expanding this relation, we can write:

\begin{displaymath}
\langle \bar{J}(\omega)\bar{J}^{*}(\omega^{\prime})\rangle = 
e^{2}\langle\sum^{N}_{k=1}\exp[\I(\omega - \omega^{\prime})t_{k}]\rangle
+ e^{2}\sum _{k\ne n}\langle\exp(\I\omega t_{k})\rangle  
\langle\exp(-\I\omega^{\prime} t_{n})\rangle \ .
\end{displaymath}

\noindent Expression $\langle\exp(\I\omega t_{k})\rangle$ is equal to 
the Fourier transformation of the bunch profile function $F(t)$:

\begin{displaymath}
\langle\exp(\I\omega t_{k})\rangle = 
\int\limits^{\infty}_{-\infty}F(t_{k})e^{\I\omega t_{k}}\D t_{k} 
= \bar{F}(\omega) \ .
\end{displaymath}

\noindent Thus we can write:

\begin{displaymath}
\langle \bar{J}(\omega)\bar{J}^{*}(\omega^{\prime})\rangle = 
e^{2}N\bar{F}(\omega - \omega^{\prime}) + 
e^2N(N-1)\bar{F}(\omega)\bar{F}(\omega^{\prime}) \ . 
\end{displaymath}

\noindent When 

\begin{equation}
N\mid\bar{F}(\omega)\mid^{2} \ll 1 \ ,
\label{eq:s2}
\end{equation}

\noindent we can write the following expression for the first-order 
spectral correlation:

\begin{equation}
\langle \bar{J}(\omega)\bar{J}^{*}(\omega^{\prime})\rangle = 
e^{2}N\bar{F}(\omega - \omega^{\prime})  \ . 
\label{eq:s3}
\end{equation}

\noindent The Fourier transform of the Gaussian profile function has 
the form:

\begin{displaymath}
\bar{F}(\omega) = \exp\left( - \frac{\omega^{2}\sigma_{\mathrm{T}}^{2}}
{2}\right) \ .
\end{displaymath}

\noindent For the specific cases of a Gaussian profile of the electron 
bunch, the first-order correlation of the complex Fourier harmonics, 
$\bar{J}(\omega)$ and $\bar{J}(\omega^{\prime})$, has the form:

\begin{equation}
\langle \bar{J}(\omega)\bar{J}^{*}(\omega^{\prime})\rangle = 
e^{2}N \exp\left[- \frac{(\omega - 
\omega^{\prime})^{2}\sigma^{2}_{\mathrm{T}}}{2}\right] \ .  
\label{eq:s4}
\end{equation}
   
Let us discuss the region of validity of the approximation 
(\ref{eq:s2}).  The physical meaning of (\ref{eq:s2}) is that the 
frequency $\omega$ has to be large enough, $\omega\sigma_{\mathrm{T}} 
\gg 1$. To be specific, we consider a numerical example for a Gaussian 
bunch profile. The value of $\mid\bar{F}(\omega)\mid^{2}$ is equal to 
$\exp(- 10^{12})$ when $\sigma_{\mathrm{T}}\omega = 10^{6}$. So 
condition (\ref{eq:s2}) is always fulfilled in practice.

\subsection{Second-order spectral correlation}

Let us calculate the second order correlation of the complex Fourier 
harmonics $\bar{J}(\omega)$ and $\bar{J}(\omega^{\prime})$:

\begin{displaymath}
\langle\mid\bar{J}(\omega)\mid^{2}\mid\bar{J}(\omega^{\prime})\mid^{2}\rangle 
= e^{4}\langle 
\sum^{N}_{n=1}\sum^{N}_{m=1}\sum^{N}_{p=1}\sum^{N}_{q=1}\exp[\I\omega 
(t_{n}-t_{m}) + \I\omega^{\prime}(t_{p}-t_{q})]\rangle \ .  
\end{displaymath}
 
\noindent The $N^{4}$ terms in this sum can be set in 15 classes 
\cite{g}.  When condition (\ref{eq:s2}) is fulfilled, only two of them 
are of importance, corresponding to $( n = m,\quad p = q, \quad n \ne p 
)$ and $( n = q, \quad m = p, \quad n \ne m )$. Thus we can write:   

\begin{equation}
\langle\mid\bar{J}(\omega)\mid^{2}\mid\bar{J}(\omega^{\prime})\mid^{2}\rangle 
= 
\langle\mid\bar{J}(\omega)\mid^{2}\rangle\langle\mid\bar{J}
(\omega^{\prime})\mid^{2}\rangle + \mid\langle\bar{J}(\omega)\bar{J}
(\omega^{\prime})\rangle\mid^{2} \ .
\label{eq:s5}
\end{equation} 

\noindent Substituting (\ref{eq:s4}) into (\ref{eq:s5}) we obtain:

\begin{equation}
\langle\mid\bar{J}(\omega)\mid^{2}\mid\bar{J}(\omega^{\prime})\mid^{2}\rangle 
= e^{4}N^{2}(1 + \mid\bar{F}(\omega - \omega^{\prime})\mid^{2}) \ .
\label{eq:s6}
\end{equation}

\subsection{The origin of shot noise in the electron beam}

Why is it that the noise in electron beam is a Gaussian random process?  
Why is this the right rule, what is the fundamental reason for it, and 
how is it connected to anything else? The explanation is deep down in 
quantum mechanics\footnote{This subsection is rather abstract side 
tour. Reader can therefore, skip over it and continue with subsection 
3.5. Noise in electron beam is a Gaussian random process. For the 
moment, reader will just have to take it as one of the  rules of the 
world .}.  It is generally accepted that quantum mechanics provides the 
best current picture of physical phenomena, and the most complete 
description of the radiation field must be sought in quantum-mechanical 
terms. The quantum theory of radiation predicts the existence of 
zero-point electromagnetic field.  The first step is to show that a 
field mode is equivalent to harmonic oscillator.  To describe a field 
mode quantum mechanically, we simply describe the equivalent harmonic 
oscillator quantum mechanically. The vacuum state $\mid 0\rangle$ has 
no photons, but predicts that it nevertheless has an energy 
$\hbar\omega/2$.  From the probability distribution for a ground-state 
harmonic oscillator, we easily obtain the probability distribution for 
the electric field in vacuum state $\mid 0\rangle$.  What is the 
physical significance of these vacuum-state expectation values? One 
thing they indicate is that the electromagnetic vacuum is a stationary 
state of the field with statistical fluctuations of the electric and 
magnetic fields. As far as measurements are concerned, however, it is 
often argued that the entire universe is evidently bathed in zero-point 
electromagnetic field, which is distributed according to a Gaussian 
probability distribution \cite{m}.

Relativistic electron beam passing through bending magnets emits 
synchrotron radiation, a process that leads to damping. 
In standard classical electrodynamics there is only the radiation reaction 
field to act on a single particle in the vacuum.  
All six degrees of freedom for electron motion in storage ring are 
damped and the actual value of electron beam emittance would be equal 
to zero.  We deduce the synchrotron radiation formula using a classical 
treatment.  So, the reasonable question arises of whether such a 
classical model describes correctly the synchrotron radiation in the 
storage ring.  Simple physical considerations show that such a 
description is  valid for any practical situation. The classical 
electrodynamics approach can be used when the energy of the radiated 
photon $\hbar\omega$ is much less than the energy of the electron.  As 
a rule , $\hbar\omega/(mc^{2}\gamma) \simeq 10^{-5}$ and quantum 
effects are negligible. Classically we generally assume implicitly that 
the homogeneous solution of the Maxwell equations is that in which the 
electric and magnetic fields vanish identically. That is, we assume 
that there are no fields in the absence of any sources. In the absence 
of sources the vacuum field is simply zero, with no energy or 
fluctuations whatsoever.  This is not to say that we cannot have 
source-free fields in classical electrodynamics. Rather, the absence of 
zero-point fields in standard classical theory lies in the assumption 
that there are no fields in the absence of any sources.

We can go beyond standard classical theory and postulate the 
existence of zero-point electric and magnetic fields in the absence of 
any sources. The difference between classical and quantum 
electrodynamics, together with the evident importance of the 
fluctuating vacuum field in quantum electrodynamics, suggests the 
adoption of a different boundary conditions in classical 
electrodynamics: instead of assuming that the classical field vanishes 
in the absence of sources, we can assume that there is a fluctuating 
classical field with zero energy $\hbar\omega/2$ per mode.  As long as 
this field satisfies the Maxwell's equations there is no a priori 
inconsistency in this assumption. The appearance of $\hbar$ in this 
modification of classical electrodynamics implies no deviation from 
conventional classical ideas, for $\hbar$ is regarded as nothing more 
than a number chosen to obtain consistency of the predictions of the 
theory with experiment.  The resulting theory is able to account 
quantum effects within a fully classical framework \cite{m}.

We note that electron storage rings have provided strong evidence for 
the reality of zero-point energy. In the experiments with stored 
electrons the zero-point energy sets a lower limit to freezing of the 
electron motion. If there were no noise associated with the vacuum 
fields, the actual value of electron beam emittance would be equal to 
zero. In storage rings damping will counteract quantum excitation 
leading to an equilibrium.  If one looks over the derivation of the 
quantum diffusion of the electrons in the storage ring, it becomes 
clear that the argument can be couched in the language of stochastic 
electrodynamics rather than quantum electrodynamics. That is , all that 
is really required in that derivation is zero-point energy 
$\hbar\omega/2$ per mode of the electromagnetic field.  Whether this 
zero-point energy is of quantum or classical origin is irrelevant for 
the purpose of deriving the diffusion coefficient - stochastic 
electrodynamics accounts perfectly well for the quantum diffusion of 
electrons in storage ring.  In this case both the field and the 
particles are treated classically and we have no practical need in 
accelerator physics for quantum electrodynamics. 
     
The diffusion terms will introduce a statistical mixing of the phases 
and after some damping times any initial azimuthal variation of the 
phase space density will be washed out.  
Electrons in a storage ring are evidently bathed in zero-point 
electromagnetic field, which is distributed according to a Gaussian 
probability distribution.       
This is the explanation of the relation between vacuum fluctuations and 
shot noise in the electron beam.

\subsection{Analysis of synchrotron radiation properties in the 
frequency domain}

Above we described the properties of the input shot 
noise signal in the frequency domain. The next step is the derivation 
of the spectral function connecting the Fourier amplitudes of the 
output field and the Fourier amplitudes of the input noise signal.  
In the first analysis of the problem, we adopt some 
rather simplifying assumptions that are only occasionally met in 
practice.  Following this simplified analysis, however, we show how the 
validity of the results can be extended to a far wider range of 
conditions than might have been thought at the start.

We will investigate the synchrotron radiation in the framework of the 
one-dimensional model.   
For simplicity we consider linear 
polarization of synchrotron radiation, and use a scalar representation 
of the radiation field. Nevertheless, all the results are valid for any 
polarization. 
The one-dimensional model describes the radiation of the plane 
electromagnetic wave

\begin{displaymath}
E_{y}(t) = \tilde{E}_{y}(t)\exp[\I\omega(z/c-t)] + {\mathrm{C.C.}} \ ,
\end{displaymath} 

\noindent by the electron beam in the planar undulator. The electric 
field of the electromagnetic wave in the time domain, $E_{y}(t)$, and 
its Fourier transform, $\bar{E}(\omega)$, are connected by

\begin{displaymath}
E_{y}(t) = 
\frac{1}{2\pi}\int\limits^{\infty}_{-\infty}\bar{E}(\omega)e^{-\I\omega 
t}\D\omega \ .
\end{displaymath} 

\noindent When $\omega < 0$ the Fourier harmonic is defined by the 
relation $\bar{E}^{*}(\omega) = \bar{E}(-\omega)$. 
The Fourier 
harmonic of the transversely coherent electromagnetic field at the 
spatial filter exit and the Fourier harmonic of the current at the 
undulator entrance are connected by the relation:

\begin{equation}
\bar{E}(\omega) = A(\omega)\bar{J}(\omega) \ , \quad 
\omega > 0 \ ,
\label{eq:e1}
\end{equation} 

\noindent where $A(\omega)$ is the spectral function of 
the undulator. For an undulator of $N_{\mathrm{w}}$ periods each 
electron oscillates through $N_{\mathrm{w}}$ cycles of its motion and 
thus radiates a wavetrain consisting of $N_{\mathrm{w}}$  
cycles of the electric field. The Fourier transform of this waveform 
which gives the spectral content of the fields, is $\sin(x)/x$, where 
$x = N_{\mathrm{w}}\pi(\Delta\omega/\omega_{0})$ and 
$\Delta\omega = (\omega - \omega_{0})$ is the frequency shift away from 
the central maximum at $\omega_{0}$. Thus we can write (see section 4): 

\begin{displaymath}
\mid A(\omega)\mid^{2} = 
A_{0}\frac{\sin^{2}(N_{\mathrm{w}}\pi\Delta\omega/\omega_{0})}
{(N_{\mathrm{w}}\pi\Delta\omega/\omega_{0})^{2}} \ .
\end{displaymath}

\noindent It is relevant to make some remarks on the region of 
applicability of the one-dimensional theory. One-dimensional model 
assumes the input shot noise and output radiation to have full 
transverse coherence. This assumption allow us to assume that the input 
shot noise signal is defined by the value of beam current 
(\ref{eq:e1}).  In reality the fluctuations of the electron beam 
current density are uncorrelated in the transverse dimension.  Using 
the notion of the beam radiation modes, we can say that many transverse 
radiation modes are radiated when the electron beam enters the 
undulator. The one-dimensional model can be used for the 
calculations of statistical properties of transversely coherent 
synchrotron radiation. In practice such an assumption is valid for 
synchrotron light at the exit of a spatial filter.  During the spatial 
filtering process, the number of transverse modes decreases, and the 
contribution of the coherent radiation into the total radiation power 
is increased up to full coherence.  
With this assumption, attention can be concentrated completely on 
temporal coherence effects. 

Here we have used the simplest model of 
synchrotron light source with zero energy spread into electron beam. In 
addition, it is assumed that the electrons move along constrained 
sinusoidal trajectories in parallel with the undulator axis.  
The first assumption is primarily a statement that the energy spread, 
$\Delta\gamma/\gamma$, is so small that 
$N_{\mathrm{w}}\Delta\gamma/\gamma \ll 1$. Such a condition is 
generally valid in practice. The second assumption is not generally 
valid in practice, but it will be removed in section 5.

Above we studied the properties of the Fourier harmonics of 
the shot noise in the frequency domain. The Fourier harmonics of the 
output radiation field are connected with the Fourier harmonics of the 
input shot noise by (\ref{eq:e1}).  It follows from (\ref{eq:e1}) that 
statistical properties of the Fourier amplitudes $\bar{E}(\omega)$ are 
defined by the statistical properties of the Fourier amplitudes of the 
input current $\bar{J}(\omega)$. In particular, it follows immediately 
from (\ref{eq:s1}) that $\mid\bar{E}(\omega)\mid^{2}$ is distributed in 
accordance with the negative exponential probability density function:  

\begin{equation}
p(\mid\bar{E}(\omega)\mid^{2}) = 
\frac{1}{\langle\mid\bar{E}(\omega)\mid^{2}\rangle} 
\exp\left(-\frac{\mid\bar{E}(\omega)\mid^{2}}{
\langle\mid\bar{E}(\omega)\mid^{2}\rangle}\right) \ . 
\label{eq:e2}
\end{equation}

For many practical applications of the synchrotron radiation a 
monochromator has to be installed at the undulator exit. We denote the 
frequency profile function of the monochromator as 
$G_{\mathrm{m}}(\omega)$.  The linearity of Maxwell's equations and the 
fact that a monochromator can be treated as a linear filter allows one 
to write the Fourier components of the electric field of the 
synchrotron radiation in the following form:

\begin{equation}
\bar{E}(\omega) = A(\omega)G_{\mathrm{m}}(\omega)
\bar{J}(\omega) \ , \qquad \omega > 0 \ ,
\label{eq:e3}
\end{equation}

Let us calculate the correlation of the complex Fourier harmonics 
$\bar{E}(\omega)$ and $\bar{E}(\omega^{\prime})$. When the 
monochromator bandwidth is much less than the spectral 
bandwidth of the undulator radiation we obtain:

\begin{equation}
\langle\bar{E}(\omega)\bar{E}^{*}(\omega^{\prime})\rangle =
\mid A(\omega_{0})\mid^{2}G_{\mathrm{m}}(\omega - \omega_{0})
G_{\mathrm{m}}^{*}(\omega^{\prime} - \omega_{0})\langle\bar{J}(\omega)
\bar{J}^{*}(\omega^{\prime})\rangle \ .
\label{eq:e4}
\end{equation} 

\noindent It is seen from (\ref{eq:s3}) that one can write 
(\ref{eq:e4}) as

\begin{equation}
\langle\bar{E}(\omega)\bar{E}^{*}(\omega^{\prime})\rangle =
e^{2}N\mid A(\omega_{0})\mid^{2}G_{\mathrm{m}}(\omega - \omega_{0})
G_{\mathrm{m}}^{*}(\omega^{\prime} - \omega_{0})\bar{F}(\omega - 
\omega^{\prime}) \ .  
\label{eq:e5}
\end{equation}

The first-order spectral correlation function is defined as

\begin{equation}
g_{1}(\omega - \omega^{\prime}) =
\frac{\langle\bar{E}(\omega)\bar{E}^{*}(\omega^{\prime})\rangle}
{\sqrt{\langle\mid\bar{E}(\omega)\mid^{2}\rangle
\langle\mid\bar{E}(\omega^{\prime})\mid^{2}\rangle}} \ .
\label{eq:e6}
\end{equation} 

\noindent Substituting (\ref{eq:e5}) into (\ref{eq:e6}), we obtain

\begin{equation}
\mid g_{1}(\omega - \omega^{\prime})\mid = \mid\bar{F}(\omega - 
\omega^{\prime})\mid \ .  
\label{eq:e7} 
\end{equation} 

\noindent The explicit expression for the first-order spectral 
correlation function of the synchrotron radiation emitted by the 
electron bunch with Gaussian profile has the form:

\begin{equation}
g_{1}(\omega - \omega^{\prime}) = 
\exp\left[-\frac{(\omega-\omega^{\prime})^{2}
\sigma_{\mathrm{T}}^{2}}{2}\right] \ . 
\label{eq:e8}
\end{equation}

We define the interval of spectral coherence as

\begin{equation}
\Delta\omega_{\mathrm{c}} = \int\limits^{\infty}_{-\infty}
\mid g_{1}(\Delta\omega)\mid^{2}\D(\Delta\omega) \ .
\label{eq:e9}
\end{equation} 

\noindent The value of the spectral coherence for the synchrotron 
source with a Gaussian electron bunch is given by

\begin{equation}
\Delta\omega_{\mathrm{c}} = \int\limits^{\infty}_{-\infty}
\mid \bar{F}(\Delta\omega)\mid^{2}\D(\Delta\omega) 
= \frac{\sqrt{\pi}}{\sigma_{\mathrm{T}}} \ .
\label{eq:e10}
\end{equation} 

The second-order spectral correlation function is defined as

\begin{equation}
g_{2}(\omega - \omega^{\prime}) =
\frac{\langle\mid\bar{E}(\omega)\mid^{2}\mid\bar
{E}(\omega^{\prime})\mid^{2}\rangle}
{\langle\mid\bar{E}(\omega)\mid^{2}\rangle
\langle\mid\bar{E}(\omega^{\prime})\mid^{2}\rangle} \ .
\label{eq:e11}
\end{equation} 

\noindent Using (\ref{eq:s6}), (\ref{eq:e3}) and (\ref{eq:e11}) we 
obtain that the first and second order correlation functions are 
connected by the relation:

\begin{equation}
g_{2}(\omega - \omega^{\prime}) = 1 + \mid g_{1}(\omega - 
\omega^{\prime})\mid^{2} \ , 
\label{eq:e12}
\end{equation} 

\noindent which is also a general property of completely chaotic 
polarized radiation. The explicit expression for the second-order 
spectral correlation function for the synchrotron source with  
Gaussian electron bunch has the form:

\begin{displaymath}
g_{2}(\omega - \omega^{\prime}) = 1 + \mid\bar{F}(\omega - 
\omega^{\prime})\mid^{2} = 1 + \exp\left[-(\omega - 
\omega^{\prime})^{2}\sigma_{\mathrm{T}}^{2}\right] \ . 
\end{displaymath}

\subsection{Energy fluctuation experiment}

In this subsection we are going to discuss the application of 
statistical optics to a practical device. We will find that many of the 
features of this specific problem are quite common in the general 
statistical theory of synchrotron light, and we will learn a great 
deal by considering this one problem in detail. 

\begin{figure}[tb]
\begin{center}
\epsfig{file=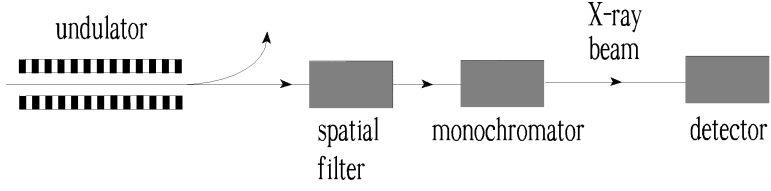,width=0.75\textwidth}
\end{center}
\caption{
Set-up of model experiment} 
\label{fig:pps9} 
\end{figure}

Figure~\ref{fig:pps9} shows the fluctuation experiment geometry under 
consideration. The problem is a 
description of the fluctuations of the energy of the radiation pulse 
$W$ at the detector installed after the spatial filter and 
monochromator.  The experiment is here analyzed in some detail to 
determine the conditions under which the field fluctuations, resulting 
from the chaotic nature of the light source, affect the energy 
fluctuations.  Note that in this experiment, both monochromator 
resolution and radiation pulse duration play a role.  There is no X-ray 
monochromator providing the coherence time $\tau_{\mathrm{c}} 
\simeq \lambda^{2}/c\Delta\lambda$ comparable with radiation pulse 
duration (100 ps). The model experiment ignores these technical 
limitations.    

Using the expression for Poynting's vector and 
Parseval's theorem, we calculate the radiation energy in one radiation 
pulse: 

\begin{displaymath}
W = \frac{cS}{4\pi}\int\limits^{\infty}_{-\infty}E_{y}^{2}(t)\D t =
\frac{cS}{4\pi^{2}}\int\limits^{\infty}_{0}\mid\bar{E}
(\omega)\mid^{2}\D\omega  \ , 
\end{displaymath} 

\noindent where $S$ is the transverse area of the detector. The 
energy, averaged over an ensemble, is given by the expression:

\begin{equation}
\langle W\rangle = \frac{cS}{4\pi^{2}}\int\limits^{\infty}_{0}
\langle\mid\bar{E}
(\omega)\mid^{2}\rangle\D\omega  \ , 
\label{eq:e13}
\end{equation} 
  
\noindent where $\langle\mid\bar{E}(\omega)\mid^{2}\rangle$ is 
calculated using (\ref{eq:s3}) and (\ref{eq:e4}):

\begin{equation}
\langle\mid\bar{E}(\omega)\mid^{2}\rangle =
\mid A(\omega)\mid^{2}\mid G_{\mathrm{m}}(\omega)
\mid^{2}\langle\mid\bar{I}(\omega)\mid^{2}\rangle =
e^{2}N\mid A(\omega)\mid^{2}\mid G_{\mathrm{m}}(\omega)\mid^{2} \ . 
\label{eq:e14}
\end{equation} 

\noindent Using (\ref{eq:e13}) and (\ref{eq:e14}), we calculate the 
average energy measured after the monochromator: 

\begin{displaymath}
\langle W\rangle = \frac{cS}{4\pi^{2}}\int\limits^{\infty}_{0}
\langle\mid\bar{E}
(\omega)\mid^{2}\rangle\D\omega = 
\frac{ce^{2}SN}{4\pi^{2}}\int\limits^{\infty}_{0}\mid 
A(\omega)\mid^{2}\mid G_{\mathrm{m}}(\omega)\mid^{2}\d\omega \ . 
\end{displaymath} 

It is seen that the average energy is a function of the frequency 
profile of the monochromator and undulator. 

The variance of the energy distribution is calculated as 
follows:

\begin{displaymath}
\sigma^{2}_{\mathrm{W}} = \frac{\langle(W-\langle W\rangle)^{2}\rangle}
{\langle W\rangle^{2}} =
\frac{\int\limits^{\infty}_{0}\D\omega
\int\limits^{\infty}_{0}\D\omega^{\prime}
\langle\mid\bar{E}(\omega)\mid^{2}\mid
\bar{E}(\omega^{\prime})\mid^{2}\rangle}
{\int\limits^{\infty}_{0}\D\omega
\langle\mid\bar{E}(\omega)\mid^{2}\rangle
\int\limits^{\infty}_{0}\D\omega^{\prime}
\langle\mid\bar{E}(\omega^{\prime})\mid^{2}\rangle}
- 1 \ .
\end{displaymath} 

\noindent Using definition (\ref{eq:e11}) of the second-order 
correlation function and (\ref{eq:e12}), we reduce this expression to

\begin{equation}
\sigma^{2}_{\mathrm{W}} = 
\frac{\int\limits^{\infty}_{0}\D\omega
\int\limits^{\infty}_{0}\D\omega^{\prime}
\langle\mid\bar{E}(\omega)\mid^{2}\rangle\langle\mid
\bar{E}(\omega^{\prime})\mid^{2}\rangle\mid g_{1}(\omega
-\omega^{\prime})\mid^{2}}
{\int\limits^{\infty}_{0}\D\omega
\langle\mid\bar{E}(\omega)\mid^{2}\rangle
\int\limits^{\infty}_{0}\D\omega^{\prime}
\langle\mid\bar{E}(\omega^{\prime})\mid^{2}\rangle} \ .
\label{eq:e15}
\end{equation} 

\noindent The analysis of this expression shows that the variance of 
the energy distribution after the monochromator is a function of the 
frequency profile of the monochromator, of the frequency profile of the 
undulator, and of the electron bunch formfactor $\bar{F}(\omega - 
\omega^{\prime})$.

Let us consider the case of an electron bunch with a Gaussian
profile and a monochromator with a rectangular line:

\begin{displaymath}
\mid G_{\mathrm{m}}(\omega - \omega_{0})\mid^{2} = 1
\qquad {\mathrm{for}} \qquad \mid\omega - \omega_{0}\mid < 
\frac{\Delta\omega_{\mathrm{m}}}{2} \ ,
\end{displaymath} 

\begin{displaymath}
\mid G_{\mathrm{m}}(\omega - \omega_{0})\mid^{2} = 0
\qquad {\mathrm{for}} \qquad \mid\omega - \omega_{0}\mid > 
\frac{\Delta\omega_{\mathrm{m}}}{2} \ .
\end{displaymath} 

\noindent For simplicity we assume that the bandwidth of the 
monochromator is small and $A(\omega)$ is constant within the 
monochromator bandwidth. Then the integration of the expression 
(\ref{eq:e15}) provides the following result:

\begin{equation}
\sigma^{2}_{\mathrm{W}} = 
\frac{\Delta\omega_{\mathrm{c}}}{\Delta\omega_{\mathrm{m}}}
{\mathrm{erf}}\left(\frac{\sqrt{\pi}\Delta\omega_{\mathrm{m}}}
{\Delta\omega_{\mathrm{c}}}\right) - \frac{1}{\pi}
\left(\frac{\Delta\omega_{\mathrm{c}}}{\Delta
\omega_{\mathrm{m}}}\right)^{2}\left\{1 - \exp\left[
-\pi\left(\frac{\Delta\omega_{\mathrm{m}}}{\Delta
\omega_{\mathrm{c}}}\right)^{2}\right]\right\} \ , 
\label{eq:e16}
\end{equation} 

\noindent where ${\mathrm{erf}}(x)$ is the error function and 
$\Delta\omega_{\mathrm{c}}$ is given by expression (\ref{eq:e10}).

Let us study the asymptotic behavior of (\ref{eq:e16}). When the 
monochromator bandwidth is much less than the interval of spectral 
coherence, the normalized dispersion tends to unity:

\begin{displaymath}
\sigma^{2}_{\mathrm{W}} \simeq 1
\qquad {\mathrm{for}} \qquad \Delta\omega_{\mathrm{m}} \ll 
\Delta\omega_{\mathrm{c}} \ .
\end{displaymath} 

\noindent When the monochromator bandwidth is much larger than the 
interval of spectral coherence, the variance is inversely proportional 
to the monochromator bandwidth:

\begin{displaymath}
\sigma^{2}_{\mathrm{W}} \simeq  \frac{\Delta\omega_{\mathrm{c}}}
{\Delta\omega_{\mathrm{m}}}
\qquad {\mathrm{for}} \qquad \Delta\omega_{\mathrm{c}} \ll 
\Delta\omega_{\mathrm{m}} \ .
\end{displaymath} 
 
The next practical problem is to find the probability density 
distribution of the radiation energy after the monochromator, $p(W)$.
Using the well-known results obtained in the framework of statistical 
optics, we can state that the distribution of the radiation energy 
after the monochromator is described rather well by the gamma 
probability density function:

\begin{equation}
p(W) = 
\frac{M^{M}}{\Gamma(M)}\left(\frac{W}{\langle{W}\rangle}\right)^{M-1} 
\frac{1}{\langle{W}\rangle}\exp\left(-M\frac{W}
{\langle{W}\rangle}\right) \ , 
\label{eq:e17}
\end{equation}

\noindent where $\Gamma(M)$ is the gamma function of argument $M$, and 
$1/M = \sigma^{2}_{\mathrm{W}}$. This distribution provides correct 
values for the mean value of $W$ and for the variance 
$\sigma^{2}_{\mathrm{W}}$:

\begin{displaymath}
\int\limits^{\infty}_{0}Wp(W)\D W = \langle W\rangle \ , \quad
\int\limits^{\infty}_{0}\frac{(W-\langle W\rangle)^{2}}
{\langle W\rangle^{2}}p(W)\D W = \frac{1}{M} \ .
\end{displaymath}

\noindent The parameter $M$ can be interpreted as the average number of 
degrees of freedom (or modes) in the radiation pulse. It follows from 
(\ref{eq:e15}) that this parameter cannot be less than unity. When $M$ 
tends to the unity, (\ref{eq:e17}) tends to the negative exponential 
distribution (\ref{eq:e2}). For large values of $M$ the distribution 
(\ref{eq:e17}) tends to a Gaussian distribution.

\subsection{Analysis of the radiation properties in the time domain}

An expression for the electric field of the electromagnetic wave as a 
function of time $t$ can be obtained using the Fourier transform of 
(\ref{eq:e3}). It is convenient to isolate explicitly the slowly varying 
complex amplitude:

\begin{equation}
E_{y}(t) = \tilde{E}(t)e^{-\I\omega_{0}t} + {\mathrm{C.C.}} \ ,      
\label{eq:e18}
\end{equation}

\noindent where

\begin{displaymath}
\tilde{E}(t)e^{-\I\omega_{0}t} = \frac{1}{2\pi}\int\limits^{\infty}_{0}
\bar{E}(\omega)e^{-\I\omega t}\D\omega \ .
\end{displaymath}

\noindent Let us consider the specific case of a 
Gaussian monochromator line profile

\begin{equation}
\mid G_{\mathrm{m}}(\omega)\mid^{2} = 
\exp\left[-\frac{(\omega-\omega_{0})^{2}}
{2\sigma^{2}_{\mathrm{m}}}\right]  
\label{eq:e19}
\end{equation}

\noindent allowing calculations of the complex amplitude $\tilde{E}(t)$ 
of the electric field at the monochromator exit in the limit of small 
monochromator bandwidth.  In other words, we assume $A(\omega)$ to be 
constant within the monochromator bandwidth.  

\noindent Using (\ref{eq:e1}), (\ref{eq:s1a}), (\ref{eq:e19}) and 
(\ref{eq:e18}), we obtain:

\begin{equation}
\tilde{E}(t) = 
E_{0}\left\{\frac{1}{N}\sum^{N}_{k=1}\exp(\I\omega_{0}t_{k})
\exp[-\sigma_{\mathrm{m}}^{2}(t-t_{k})^{2}]\right\} \ .
\label{eq:e20}
\end{equation}

\noindent Since $\sigma_{\mathrm{m}} \ll \omega_{0}$, we can 
approximately let the amplitude and the phase of each random phasor 
contributing to the sum in (\ref{eq:e20}) be independent of each other, 
and the phases be uniformly distributed on the interval $(0,2\pi)$. 
Thus, the distribution of the instantaneous radiation power density $(\ 
\propto \ \mid\tilde{E}(t)\mid^{2})$ is the negative exponential 
distribution:  

\begin{equation}
p(\mid\tilde{E}(t)\mid^{2}) = 
\frac{1}{\langle\mid\tilde{E}(t)\mid^{2}\rangle} 
\exp\left(-\frac{\mid\tilde{E}(t)\mid^{2}}{
\langle\mid\tilde{E}(t)\mid^{2}\rangle}\right) \ . 
\label{eq:e2a}
\end{equation}

The correlation between the radiation fields at times $t$ and 
$t^{\prime}$ has the form:

\begin{eqnarray}
& \mbox{} &
\langle\tilde{E}(t)\tilde{E}^{*}(t^{\prime})\rangle 
\exp[-\I\omega_{0}(t-t^{\prime})]
\nonumber\\
& \mbox{} &
= 
\frac{1}{4\pi^{2}}\int\limits^{\infty}_{0}\D\omega\int\limits^{\infty}_{0}
\D\omega^{\prime}\exp(-\I\omega t + \I\omega^{\prime}t^{\prime})
\langle\bar{E}(\omega)\bar{E}^{*}(\omega^{\prime})\rangle
\nonumber\\
& \mbox{} &
= \frac{1}{4\pi^{2}}\int\limits^{\infty}_{0}\D\omega\int\limits^{\infty}_{0}
\D\omega^{\prime}\exp(-\I\omega t + \I\omega^{\prime}t^{\prime})
\nonumber\\
& \mbox{} &
\times
[\langle\mid\bar{E}(\omega)\mid^{2}\rangle
\langle\mid\bar{E}(\omega^{\prime})\mid^{2}\rangle]^{1/2}
\bar{F}(\omega-\omega^{\prime}) \ .
\label{eqn1}
\end{eqnarray}

\noindent Here (\ref{eq:e6}) and (\ref{eq:e7}) have been used when 
rewriting the integral.  In the present consideration we use 
the approximation which assumes the interval of spectral coherence  
is much less than the monochromator bandwidth 

\begin{equation}
\Delta\omega_{\mathrm{c}}  \ll \sigma_{\mathrm{m}} \ .
\label{eq:e1a}
\end{equation}

\noindent Thus we can simplify the integral in 
the following way. We replace the expression in square brackets 
by $\langle\mid\bar{E}(\omega)\mid^{2}\rangle$, and after integration 
over $\omega$ and $\Delta\omega = (\omega-\omega^{\prime})$ we obtain:

\begin{eqnarray}
& \mbox{} &
\langle\tilde{E}(t)\tilde{E}^{*}(t^{\prime})\rangle 
\exp[-\I\omega_{0}(t-t^{\prime})]
\nonumber\\
& \mbox{} &
= 
\frac{1}{4\pi^{2}}\int\limits^{\infty}_{0}\D(\Delta\omega)
\exp(-\I\Delta\omega t)\bar{F}(\Delta\omega)
\int\limits^{\infty}_{0}
\D\omega\exp[-\I\omega(t-t^{\prime})]\langle
\mid\bar{E}(\omega)\mid^{2}\rangle
\nonumber\\
& \mbox{} &
= \frac{F(t)}{2\pi}\int\limits^{\infty}_{0}
\D\omega\exp[-\I\omega(t-t^{\prime})]\langle
\mid\bar{E}(\omega)\mid^{2}\rangle \ ,
\label{eqn2}
\end{eqnarray}  

\noindent where $F(t)$ is the radiation pulse profile. We define the 
first-order time correlation function as follows:

\begin{displaymath}
g_{1}(t-t^{\prime}) =
\frac{\langle\tilde{E}(t)\tilde{E}^{*}(t^{\prime})\rangle}
{\left[\langle\mid\tilde{E}(t)\mid^{2}\rangle
\langle\mid\tilde{E}(t^{\prime})\mid^{2}\rangle\right]^{1/2}} \ .
\end{displaymath} 

\noindent Using (\ref{eqn2}), we can write  
        
\begin{displaymath}
g_{1}(t-t^{\prime}) =
\frac{\int\limits^{\infty}_{0}\D\omega\langle\mid\bar{E}(\omega)
\mid^{2}\rangle\exp[-\I(\omega-\omega_{0})(t-t^{\prime})]}
{\int\limits^{\infty}_{0}\D\omega\langle
\mid\bar{E}(\omega)\mid^{2}\rangle}
\end{displaymath}

\noindent Since we deal with a narrow-band signal, the latter 
expression may be rewritten as follows

\begin{equation}
g_{1}(t-t^{\prime}) =
\frac{\int\limits^{\infty}_{-\infty}\D(\Delta\omega)
\langle\mid\bar{E}(\Delta\omega)
\mid^{2}\rangle\exp[-\I(\Delta\omega)(t-t^{\prime})]}
{\int\limits^{\infty}_{-\infty}\D(\Delta\omega)\langle
\mid\bar{E}(\Delta\omega)\mid^{2}\rangle} \ ,
\label{eq:e21}
\end{equation}

\noindent where $\Delta\omega = (\omega-\omega_{0})$. Therefore, the 
slowly varying correlation function and the normalized spectrum of the 
narrow-band signal are a Fourier transform pair. Remembering relation 
(\ref{eq:e1}), we rewrite (\ref{eq:e21}) in the following way:

\begin{displaymath}
g_{1}(t-t^{\prime}) =
\frac{\int\limits^{\infty}_{-\infty}\D(\Delta\omega)
\mid G_{\mathrm{m}}(\Delta\omega)
\mid^{2}\exp[-\I(\Delta\omega)(t-t^{\prime})]}
{\int\limits^{\infty}_{-\infty}\D(\Delta\omega)
\mid G_{\mathrm{m}}(\Delta\omega)\mid^{2}} \ .
\end{displaymath}

\noindent It is seen from this expression that the first-order 
correlation function possesses the property $g_{1}(t-t^{\prime}) =
g_{1}^{*}(t^{\prime}-t)$. 
When the monochromator line profile is symmetrical 
the function $g_{1}$ is real. For instance, for a monochromator with 
Gaussian line profile the correlation function $g_{1}$ is

\begin{equation}
g_{1}(\tau) = 
\exp\left(-\frac{\sigma^{2}_{\mathrm{m}}\tau^{2}}{2}\right) \ , 
\label{eq:e22}
\end{equation}

\noindent where $\tau = (t-t^{\prime})$. Following the approach of 
Mandel, we define the coherence time, $\tau_{\mathrm{c}}$, as

\begin{displaymath}
\tau_{\mathrm{c}} = \int\limits^{\infty}_{-\infty}\mid g_{1}
(\tau)\mid^{2}\D\tau \ .
\end{displaymath}

\noindent For the case of the Gaussian line profile , the explicit 
expression for the coherence time is

\begin{equation}
\tau_{\mathrm{c}} = \frac{\sqrt{\pi}}{\sigma_{\mathrm{m}}} \ .
\label{eq:e23}
\end{equation}

The correlation between the radiation intensities at times $t$ and 
$t^{\prime}$ is defined as

\begin{eqnarray}
& \mbox{} &
\langle\mid\tilde{E}(t)\mid^{2}\mid\tilde{E}(t^{\prime})\mid^{2}\rangle 
\nonumber\\
& \mbox{} &
= 
\frac{1}{16\pi^{4}}\int\limits^{\infty}_{0}\D\omega_{1}
\int\limits^{\infty}_{0}\D\omega_{2}
\int\limits^{\infty}_{0}\D\omega_{3}
\int\limits^{\infty}_{0}\D\omega_{4}
\exp[-\I(\omega_{1}-\omega_{2})t - 
\I(\omega_{3}-\omega_{4})t^{\prime})]
\nonumber\\
& \mbox{} &
\times 
\langle\bar{E}(\omega_{1})\bar{E}(\omega_{3})
\bar{E}^{*}(\omega_{2})\bar{E}^{*}(\omega_{4})\rangle \ . 
\label{eqn3}
\end{eqnarray}  

\noindent Taking into account (\ref{eq:e3}), (\ref{eq:s1a}) and 
(\ref{eq:s2}), we simplify the correlation in the integrand of 
(\ref{eqn3}) in the following way:

\begin{eqnarray}
& \mbox{} & 
\langle\bar{E}(\omega_{1})\bar{E}(\omega_{3})
\bar{E}^{*}(\omega_{2})\bar{E}^{*}(\omega_{4})\rangle  
\nonumber\\
& \mbox{} &
= 
\langle\bar{E}(\omega_{1})\bar{E}^{*}(\omega_{2})\rangle
\langle\bar{E}(\omega_{3})\bar{E}^{*}(\omega_{4})\rangle  
\nonumber\\
& \mbox{} &
+ 
\langle\bar{E}(\omega_{1})\bar{E}^{*}(\omega_{4})\rangle
\langle\bar{E}(\omega_{3})\bar{E}^{*}(\omega_{2})\rangle \ ,  
\label{eqn4}
\end{eqnarray}  

\noindent and present the integral (\ref{eqn3}) as the sum of two 
terms.  To calculate the integral, one should take into account 
(\ref{eq:e6}), (\ref{eq:e7}), (\ref{eqn2}) and (\ref{eq:e1a}) , which 
leads to

\begin{equation}
\langle\mid\tilde{E}(t)\mid^{2}\mid\tilde{E}(t^{\prime})\mid^{2}\rangle
= \langle\mid\tilde{E}(t)\mid^{2}\rangle
\langle\mid\tilde{E}(t^{\prime})\mid^{2}\rangle
+ \mid\langle\tilde{E}(t)\tilde{E}^{*}(t^{\prime})\rangle\mid^{2} \ .
\label{eq:e24}
\end{equation}

The second-order time correlation function is defined as follows:

\begin{equation}
g_{2}(t-t^{\prime}) =
\frac{\langle\mid\tilde{E}(t)\mid^{2}\mid\tilde{E}
(t^{\prime})\mid^{2}\rangle}
{\langle\mid\tilde{E}(t)\mid^{2}\rangle
\langle\mid\tilde{E}(t^{\prime})\mid^{2}\rangle} \ .
\label{eq:e25}
\end{equation} 

\noindent It follows from (\ref{eq:e24}) and definitions of 
$g_{1}(t-t^{\prime})$ and $g_{2}(t-t^{\prime})$ that

\begin{equation}
g_{2}(t-t^{\prime}) = 1 + \mid g_{1}(t-t^{\prime})\mid^{2} \ .
\label{eq:e26}
\end{equation} 

\noindent The analysis of the obtained relations (\ref{eq:e2a}) and 
(\ref{eq:e26}) shows that the synchrotron radiation possesses all the 
features of completely chaotic polarized light (see also (\ref{eq:e2}) 
and (\ref{eq:e12}) obtained in the frequency domain).

\subsection{Statistical properties of the integrated intensity}

In a variety of  
problems, related with photon-counting statistics, finite-time 
integrals of instantaneous intensity occur. In many cases the 
fluctuations in the intensity are too rapid for direct observation, 
and what is measured is some average of the fluctuations over detector 
response time. The calculation can be performed without great 
difficulty in two limiting cases, namely, the response time very long 
and very short compared with the electron pulse duration. In section 
3.6 we discussed in detail fluctuations of the energy of the radiation 
pulse at the slow detector installed after the monochromator (see 
Fig.~\ref{fig:pps8}). Note that for this class of experiments we 
actually assume that the integration time is much longer than the 
electron pulse duration.  

Next let us consider the opposite extreme.
For our simplified discussion here let us assume that 
the integration time is much shorter than the electron pulse duration, 
leaving the electron pulse envelope unaffected. There is no 
photodetector providing a time resolution  much shorter than 
the X-ray pulse duration.  Present-day X-ray detection will be limited 
in time resolution of about 100 ps. The situation is somewhat different 
in pump-probe experiments.  A pump-probe experiment has analogies with 
a femtosecond photodetector measurement. The basic analogy is that a 
femtosecond pump pulse interacts with the sample and excites it into 
non-equilibrium state. The sample thereafter relaxes towards a new 
equilibrium state.  Typical relaxation times are in the femtosecond 
range. This process can be mapped by sending a X-ray probe pulse onto 
the sample.  Such a device could be discussed as a femtosecond X-ray 
detector by recording a secondary process such as photoelectrons.  

Several assumptions are made about the profile of the photodetector 
gating function.  For the present, consideration is restricted to the 
simplest case from the analytical point of view, namely, completely 
chaotic polarized radiation and an ideal detector with a stepped 
profile gating function.  So, the next problem is the description of 
the fluctuations of the radiation energy $W$ detected during a finite 
time interval $\delta T$:

\begin{displaymath}
W = \int\limits^{t+\delta T}_{t}I(t)\D t \ . 
\end{displaymath} 

\noindent The variance is calculated as follows:

\begin{eqnarray}
& \mbox{} & 
\sigma^{2}_{\mathrm{W}} = \frac{\langle(W-\langle W\rangle)^{2}\rangle}
{\langle W\rangle^{2}} =
\frac{1}{(\delta T)^{2}}
\int\limits^{t+\delta T}_{t}\D t^{\prime}
\int\limits^{t+\delta T}_{t}\D t^{\prime\prime}
\frac{\langle I(t^{\prime})I(t^{\prime\prime})\rangle}{\langle 
I\rangle^{2}} - 1 
\nonumber\\
& \mbox{} &
= 
\frac{1}{(\delta T)^{2}}
\int\limits^{t+\delta T}_{t}\D t^{\prime}
\int\limits^{t+\delta T}_{t}\D t^{\prime\prime}
\frac{\langle\mid\tilde{E}(t^{\prime})\mid^{2}\mid\tilde{E}
(t^{\prime\prime})\mid^{2}\rangle}
{\langle\mid\tilde{E}(t^{\prime})\mid^{2}\rangle
\langle\mid\tilde{E}(t^{\prime\prime})\mid^{2}\rangle} - 1 
\nonumber\\
& \mbox{} &
=
\frac{1}{(\delta T)^{2}}
\int\limits^{\delta T}_{0}\D u
\int\limits^{\delta T}_{0}\D v
g_{2}(u-v) - 1  \ .
\label{eqn5}
\end{eqnarray}  
 
\noindent Substituting (\ref{eq:e25}), (\ref{eq:e26}) and 
(\ref{eq:e22}) into (\ref{eqn5}) we obtain the following expression for 
the variance of the energy distribution detected during time interval 
$\delta T$ for the Gaussian monochromator line profile:

\begin{eqnarray}
& \mbox{} & 
\sigma^{2}_{\mathrm{W}} = 
\frac{1}{(\delta T)^{2}}
\int\limits^{\delta T}_{0}\D u
\int\limits^{\delta T}_{0}\D v
\mid g_{1}(u-v)\mid^{2}   
\nonumber\\
& \mbox{} &
=
\frac{1}{(\delta T)^{2}}
\int\limits^{\delta T}_{-\delta T}(\delta T - \tau)
\mid g_{1}(\tau)\mid^{2}\D\tau
\nonumber\\
& \mbox{} &
=
\frac{\sqrt{\pi}}{\delta T\sigma_{\mathrm{m}}}{\mathrm{erf}}
(\delta T\sigma_{\mathrm{m}}) -
\frac{1 - \exp[-(\delta T\sigma_{\mathrm{m}})^{2}]}
{(\delta T\sigma_{\mathrm{m}})^{2}} \ .
\label{eqn6}
\end{eqnarray}  

\noindent It can be shown that the integrated intensity distribution is 
described rather well by the gamma probability density function 
(\ref{eq:e12}) with parameter $M$ equal to $\sigma_{\mathrm{W}}^{-2}$. 
When $\delta T$ is less than the coherence time $\tau_{\mathrm{c}}$ 
(\ref{eq:e23}), the parameter $M$ tends to unity and the gamma 
distribution tends to the negative exponential distribution. In the 
opposite case, when $\delta T \gg \tau_{\mathrm{c}}$, we can write:

\begin{displaymath}
M^{-1} = \frac{\sqrt{\pi}}{\delta T\sigma_{\mathrm{m}}} =
\frac{\tau_{\mathrm{c}}}{\delta T} \ ,
\end{displaymath} 

\noindent and the gamma distribution tends to a Gaussian distribution.

In some experiments, such as one in which a long probe pulse is being 
used to measure lifetime of internal conversion of 
electronic excitation, the photocount time dependence is exponential 
dependence. In this case the detector gating function is an exponential 
function and the variance is calculated as follows:

\begin{figure}[tb]
\begin{center}
\epsfig{file=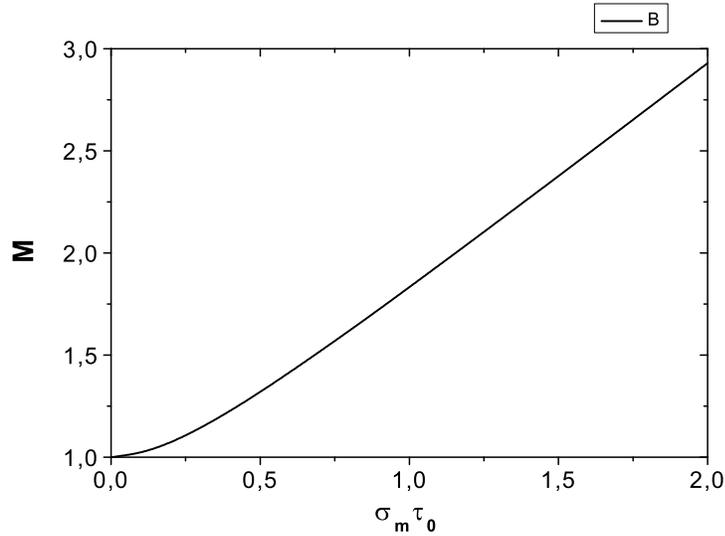,width=0.75\textwidth}
\end{center}
\caption{
Plot of $M = \sigma^{-2}_{\mathrm{W}}$ versus 
$\sigma_{\mathrm{m}}\tau_{0}$, exact solution for Gaussian spectral 
profile and exponential dependence of gating function} 
\label{fig:Graph1b} 
\end{figure}

\begin{equation}  
\sigma^{2}_{\mathrm{W}} = 
\frac{\int\limits^{\infty}_{0}\D 
t\int\limits^{\infty}_{0}\D 
t^{\prime}\exp(-t/\tau_{0})\exp(-t^{\prime}/\tau_{0})
\langle\mid\tilde{E}(t)\mid^{2}
\mid\tilde{E} (t^{\prime})\mid^{2}\rangle}
{\int\limits^{\infty}_{0}\D 
t\exp(-t/\tau_{0})\langle\mid\tilde{E}(t)\mid^{2}\rangle 
\int\limits^{\infty}_{0}\D 
t^{\prime}\exp(-t^{\prime}/\tau_{0})
\langle\mid\tilde{E}(t^{\prime})\mid^{2}\rangle} - 1 \ .  
\label{eq:e27} 
\end{equation} 

\noindent Let the monochromator line profile be Gaussian.
Substituting (\ref{eq:e25}), (\ref{eq:e26}) and 
(\ref{eq:e22}) into (\ref{eq:e27}) we obtain an expression for the 
variance of the energy distribution:

\begin{equation}
\sigma^{2}_{\mathrm{W}} = 
\int\limits^{\infty}_{0}\D u \exp(-u)
\int\limits^{\infty}_{0}\D v\exp(-v)
\exp[-\sigma^{2}_{\mathrm{m}}\tau^{2}_{0}(u-v)^{2}]   \ .
\label{eq:e28}
\end{equation}

\noindent There is no simple analytical expression for the variance in 
this case. In Fig. \ref{fig:Graph1b} we present the results of 
numerical calculations.

Our discussion of integrated intensity 
fluctuations would be incomplete if we did not refer the reader to 
another method of calculation. The sample dynamics is in the 
femtosecond range and $\langle I\rangle = {\mathrm{const.}}$ is a good 
assumption.  Clearly, we can assume that the synchrotron light in 
question is adequately modelled as an ergodic and hence stationary 
random process.  Surely the preferred way to solve the problem of 
integrated intensity fluctuations must be stationary random process 
way. Although the origins of the above viewpoint are quite clear and 
understandable, the conclusions reached regarding the relative merits 
of stationary and nonstationary analysis are very greatly in error, for 
several important reasons.

The second method of calculation has the advantage of conceptual 
simplicity but the disadvantage of being not quite as general as the 
first method. In defence of our nonstationary analysis we must say that 
models constructed for nonstationary analysis are inherently more 
general and flexible.  Indeed, they invariably contain the model of 
a stationary random process as a special case.  In the normal course of 
events, a physicist first encounters statistical optics in an entirely 
stationary random process framework.  In this case correlation 
functions in the frequency domain are represented by delta functions.  
The physicists emerging from such an introductory course may feel 
confident that they have grasped the basic statistical optics concepts 
and are ready to find a precise answer to almost any problem that comes 
their way.  Nevertheless, there is no difficulty in compiling a long 
list of examples for which an analysis of nonstationary random 
processes is required. 
     
The nonstationary random process approach is needed somewhat more 
complex than the stationary random process approach, for it requires 
knowledge of the statistical properties of synchrotron radiation not 
only in the time domain but also in the frequency domain. In the long 
run, however, the nonstationary random process model is far more 
powerful and useful than the stationary random process model in solving 
physical problems of practical interest.

\section{Photocount fluctuations in the detector of synchrotron 
radiation}

Almost all devices that measure the intensity of a light beam depend 
for their operation on the absorption of a portion of the beam, whose 
energy is converted to a detectable form. The photographic plate, 
the photomultiplier tube, and ionization and detection of 
photoelectrons in pump-probe experiments all fall into this 
category. The present section therefore considers the absorption of 
light in more detail in order to understand the nature of the intensity 
measurement process. It is convenient to base the discussion on a 
practical intensity-measuring device, and we select the photodetector. 
The photodetector depends for its operation on the 
photoelectric effect.  Since the vast majority of pump-probe detection 
problems do indeed rest on the photoelectric effect, there is 
relatively little loss of generality by making this assumption. 

Consider a storage ring synchrotron light source. The light 
from this source falls on the photosurface, and we wish to determine 
the statistical distribution of the number of photoevents observed in 
any radiation pulse. Any measurement of light 
will be accompanied by certain unavoidable fluctuations. We attribute 
these fluctuations to quantum effects; that is, light can be absorbed 
only in quanta. We shall deal with the so-called semiclassical theory 
of photodetection. Such an approach has the benefit of being 
comparatively simple in terms of the mathematical background required, 
as well as allowing a greater use of physical intuition. The 
distinguishing characteristic of this formalism is the fact that 
electromagnetic fields are treated in a completely classical manner 
until they interact with the atoms of the photosensitive material on 
which they are incident. Thus there is no necessity to deal with 
quantization of the electromagnetic field; only the interaction of the 
classical field and matter is quantized.  Fortunately it has been shown 
that the predictions of the semiclassical theory are in complete 
agreement wit the predictions of the more rigorous quantum mechanical 
approach for all detection problems involving the photoelectric effect 
\cite{g}.

We assume that the radiation reaching the detector has full 
transverse coherence and that the statistical properties of the 
radiation follow the laws described in the previous section. It has 
been shown above that the energy, $W$, in the radiation pulse is 
unpredictable.  Thus, we can predict the probability density $p(W)$ 
only. In this case the probability of detection of $K$ photons is given 
by Mandel's semiclassical formula \cite{g}:

\begin{equation}
P(K) = \int\limits^{\infty}_{0}\frac{(\alpha W)^{K}}{K!}\exp(-\alpha 
W)P(W)\D W \ ,
\label{eq:m}
\end{equation}

\noindent where $\alpha = \eta/\hbar\omega_{0}$ and $\eta$ is the 
quantum efficiency of the photodetector. It should be noted that, when 
conditioned by knowledge of energy $W$, the number of counts $K$ is a 
Poisson variable with mean $\alpha W$. Using formula (\ref{eq:m}) 
we get the expression for the mean and for the variance of the value of 
$K$:

\begin{equation}
\langle K\rangle = \alpha\langle W\rangle \ , \quad 
\sigma^{2}_{\mathrm{K}} = \frac{\langle K^2\rangle - \langle 
K\rangle^{2}}{\langle K\rangle^{2}} = \frac{1}{\langle K\rangle} + 
\sigma^{2}_{\mathrm{W}} \ , 
\label{eq:m1}
\end{equation}
                      
\noindent where $\sigma^{2}_{\mathrm{W}} = 1/M$ is given by the 
expression (\ref{eq:e15}). The expression for photocount fluctuations 
contains two terms. The first term corresponds to the photon "shot 
noise" and its origin is in the Poisson distribution. The second term 
corresponds to the classical fluctuations of the energy in the 
radiation bunch and takes its origin from the shot noise in the 
electron bunch. The ratio of the classical variance to the "photon shot 
noise" variance is named as a photocount degeneracy parameter 
$\delta_{\mathrm{c}}$:
     
\begin{equation}
\delta_{\mathrm{c}} = \frac{\langle K\rangle}{M} \ .
\label{eq:m3}
\end{equation}

The probability density of the  integrated synchrotron radiation 
intensity, $p(W)$, is the gamma distribution.  
Substituting (\ref{eq:e17}) into Mandel's formula (\ref{eq:m}) and 
performing integration we come to the negative binomial distribution 
\cite{g}:

\begin{equation}   
P(K) = \frac{\Gamma(K + M)}{\Gamma(K + 1)\Gamma(M)}\left(1 + 
\frac{M}{\langle K\rangle}\right)^{-K}\left(1 + \frac{\langle 
K\rangle}{M}\right)^{-M} \ . 
\label{eq:n}
\end{equation}

\noindent When the monochromator has a narrow linewidth, 
$\Delta\omega_{\mathrm{m}}\tau_{0} \ll 1$, parameter 
$M$ tends to the unity, and the negative binomial distribution 
transforms to the Bose distribution:

\begin{displaymath}   
\lim_{M\to 1}P(K) = \frac{\langle K\rangle^{K}}{(1 + \langle 
K\rangle)^{K+1}} \ . 
\end{displaymath}

\noindent The negative binomial distribution tends to the gamma 
distribution at large values of the count degeneracy parameter 
$\delta_{\mathrm{c}}$. In particular, the Bose distribution tends to 
the negative exponential distribution ($\delta_{\mathrm{c}} \simeq 
\langle K\rangle$ in this case):

\begin{equation}   
\lim_{\langle K\rangle\to \infty}\frac{\langle K\rangle^{K}}{(1 + 
\langle K\rangle)^{K+1}} = \frac{1}{\langle 
K\rangle}\exp\left(-\frac{K}{\langle K\rangle}\right) \ .  
\label{eq:ne}
\end{equation}
 
\noindent In the opposite case, at $\delta_{\mathrm{c}} \to 0$, the 
negative binomial distribution (\ref{eq:n}) transforms to the Poisson 
distribution:

\begin{displaymath}   
\lim_{\delta_{\mathrm{c}}\to 0}P(K) = \frac{\langle 
K\rangle^{K}}{K!}\exp(-\langle K\rangle) \ .  
\end{displaymath}

\section{Characteristics of undulator radiation}

For the proposed concept of time-resolved experiments based on the 
statistical properties, the critical parameter determining the 
performance is the count degeneracy parameter, or the average number of 
photoevents produced in a single coherence interval of the incident 
radiation. The important role of this parameter is emphasized by giving 
it a name of its own.  Because the count degeneracy parameter is 
proportional to $\langle K\rangle$, it is also proportional to the 
quantum efficiency of the photosurface.  Sometimes it is useful to 
remove this dependence on the characteristics of the particular 
detector that may be present, and to deal with a degeneracy parameter 
that is a property of the synchrotron light itself.  We thus define the 
wave degeneracy parameter as $\delta_{\mathrm{W}} = 
\delta_{\mathrm{c}}/\eta$. This new degeneracy parameter may be 
considered to be the count degeneracy parameter that would be obtained 
with an ideal detector having a quantum efficiency of unity. The wave 
degeneracy parameter is simply the average number of photons per mode.
  
In this section we are particularly concerned about the wave degeneracy 
parameter. From this section we will see that the 
degeneracy parameter is $\delta_{\mathrm{W}} \simeq 
\lambda^{3}B_{\mathrm{peak}}/(4c)$, where $B_{\mathrm{peak}}$ is the 
peak spectral brightness of the radiation source. Third-generation 
synchrotron light sources achieve high spectral brightness through very 
low-emittance electron beams and by inserting into the electron beam a 
device known as an undulator. The undulator causes the electron beam to 
undulate or bend back and forth, and this produces very bright and 
coherent light.

\subsection{Number of photons radiated in the central radiation cone}

Let us introduce the basic features of undulator radiation. 
The undulator equation 

\begin{displaymath}
\omega = 2ck_{\mathrm{w}}\gamma^{2}\left[1 + 
\frac{K_{\mathrm{w}}^{2}}{2} + \gamma^{2}\theta^{2}\right]^{-1} 
\end{displaymath}

\noindent tells us the frequency of radiation as a function of 
undulator period $\lambda_{\mathrm{w}} = 2\pi k_{\mathrm{w}}$, 
undulator parameter $K_{\mathrm{w}}$, electron energy 
$m_{0}c^{2}\gamma$, and polar angle of observation $\theta$. Note that 
for radiation within the cone of half angle

\begin{displaymath}
\theta_{\mathrm{cen}} = \frac{\sqrt{1 + 
K_{\mathrm{w}}^{2}/2}}{\gamma\sqrt{N_{\mathrm{w}}}} \ , 
\end{displaymath}

\noindent the relative spectral FWHM bandwidth is $\Delta\omega/\omega 
= 0.88/N_{\mathrm{w}}$, where $N_{\mathrm{w}}$ is the number of 
undulator periods. The spectral and angular density of the radiation 
energy emitted by a single electron during the undulator pass is given 
by the expression (at zero angle):

\begin{displaymath}
\frac{\D^{2}{\cal E}}{\D\omega\D\Omega} =
\frac{e^{2}N_{\mathrm{w}}\gamma^{2}A^{2}_{\mathrm{JJ}}K_{\mathrm{w}}^{2}}
{2c(1 + K_{\mathrm{w}}^{2}/2)^{2}}\frac{\sin^{2}[\pi 
N_{\mathrm{w}}(\omega-\omega_{0})/\omega_{0}]}
{[\pi N_{\mathrm{w}}(\omega-\omega_{0})/\omega_{0}]^{2}} \ . 
\end{displaymath}

\noindent Here $\omega_{0} = 2\gamma^{2}k_{\mathrm{w}}/(1 + 
K_{\mathrm{w}}^{2}/2)$ is the resonance frequency, $A_{\mathrm{JJ}} = 
[J_{0}(Q) - J_{1}(Q)]$, where $J_{n}$ is the Bessel function of $n$th 
order, and $Q = K_{\mathrm{w}}^{2}/(4 + 2K_{\mathrm{w}}^{2})$. Now we 
would like to calculate the energy radiated into the central cone.  In 
the small-angle approximation the solid angle is equal to $\D\Omega = 
\theta\D\theta\D\varphi$.  Integration of spectral and angular density 
over $\omega$ and $\varphi$ gives us factors 
$\omega_{0}/N_{\mathrm{w}}$ and $2\pi$, respectively.  We also have to 
integrate over $\theta$ from $0$ to $\theta_{\mathrm{cen}}$. Thus, the 
energy radiated into the central cone by a single electron is given by 

\begin{displaymath}
\Delta{\cal E}_{\mathrm{cen}} \simeq \frac{\pi 
e^{2}A^{2}_{\mathrm{JJ}}\omega_{0} K_{\mathrm{w}}^{2}}{c(1 + 
K_{\mathrm{w}}^{2}/2)} \ .  
\end{displaymath}

\noindent Individual positions of electrons in the bunch are random, 
thus the radiated fields 
due to different electrons are uncorrelated, and the average energy 
radiated by the bunch is a simple sum of the radiated energy from the 
individual electrons.    

Beyond the natural broadening, due to the finite number $N_{\mathrm{w}}$ 
of oscillations, further spectral broadening can be incurred with the 
passage of many electrons through the undulator in a bunch of finite 
size, divergence, and energy spread \cite{a}.  
If there is an electron energy (rms) spread within the bunch, 
$\Delta\gamma/\gamma$, there will be a corresponding photon energy 
spread given by $\Delta E_{\mathrm{ph}}/E_{\mathrm{ph}} = 
2\Delta \gamma/\gamma$, where the factor of two is due to the square 
relationship between photon energy and electron energy. 
A more significant effect is that due to angular distribution within 
the bunch. As a result, some electrons traverse the undulator not along 
or parallel to the $z$-axis, but at a small angle $\theta_{x,y}$. These 
electrons undergo the same number $N_{\mathrm{w}}$ of oscillations, but 
experience a somewhat longer period 
$\lambda_{\mathrm{w}}/\cos(\theta_{x,y})$. If there is an rms angular 
divergence $\sigma^{\prime}$ within the bunch, there will be a 
corresponding photon energy spread given by
$\Delta E_{\mathrm{ph}}/E_{\mathrm{ph}} = 
\gamma^{2}(\sigma^{\prime})^{2}/(1 + K_{\mathrm{w}}^{2}/2)$.
In what follows 
we use the following assumption:

\begin{equation}
4\frac{\Delta \gamma}{\gamma} \ll \frac{1}{N_{\mathrm{w}}} \ , \quad
(\sigma^{\prime})^{2} \ll \frac{1 + K_{\mathrm{w}}^{2}/2}
{\gamma^{2}N_{\mathrm{w}}} \ .  
\label{con} 
\end{equation}

\noindent When these conditions are satisfied, the energy spread and 
angular divergence cause a spectral broadening less than 
$1/N_{\mathrm{w}}$ and the central cone will be rather well defined in 
terms of both its angular definition and spectrum.

To complete the calculation of the undulator characteristics 
an expression is needed for the photon flux within the central 
cone.  We divide the radiated energy by the energy per 
photon $\hbar\omega_{0}$, and obtain

\begin{displaymath}
\frac{\D N_{\mathrm{cen}}}{\D t} = \frac{\pi\alpha A^{2}_{\mathrm{JJ}}
K_{\mathrm{w}}^{2}N f}{1 + K_{\mathrm{w}}^{2}/2} \ ,
\end{displaymath}

\noindent
where $\alpha \simeq 1/137$ is fine-structure constant, 
$N$ is the number of electrons in a bunch, $f$ is the 
bunch repetition rate.
 
\subsection{Spectral brightness of undulator radiation}

The quality of the radiation source is described usually by the 
spectral brightness defined as the density of photons in the 
six-dimensional phase space volume \cite{w}: 

\begin{displaymath}
B = 
\frac{1}{4\pi^{2}\sigma_{x}\sigma_{x}^{\prime}
\sigma_{y}\sigma_{y}^{\prime}}
\left(\lambda\frac{\D^{2}N_{\mathrm{ph}}}{\D\lambda\D t}\right) \ ,
\end{displaymath}
 
\noindent where $\sigma_{x,y}$ and $(\sigma_{x,y})^{\prime}$ are the 
(horizontal and vertical) photon beam Gaussian ($1/\sqrt{e}$) radius 
and Gaussian half angle respectively.  In the case of finite particle 
beam emittance the photon brightness is reduced. The amount of 
reduction, however, depends on the matching to the photon beam. The 
particle beam parameters are

\begin{displaymath}
(\sigma_{\mathrm{e}})^{2}_{x,y} = 
(\epsilon_{\mathrm{e}})_{x,y}\beta_{x,y} \ , \quad 
(\sigma^{\prime}_{\mathrm{e}})^{2}_{x,y} = 
(\epsilon_{\mathrm{e}})_{x,y}/\beta_{x,y} \ , 
\end{displaymath}

\noindent where $\beta_{x,y}$ is the betatron function in the 
undulator, $\epsilon_{x,y}$ is the horizontal and vertical electron 
beam emittance, respectively.  
To modify the expression for spectral brightness for the case of 
undulator radiation, we can use the previously calculated photon flux 
in the central radiation cone, $\D N_{\mathrm{cen}}/\D t$, which was 
defined as having a relative spectral bandwidth (BW) of 
$\Delta\omega/\omega_{0} = 1/N_{\mathrm{w}}$ (under condition 
(\ref{con})) and radiation cone of half angle $\theta_{\mathrm{cen}}$. 
Taking into account the contribution of diffraction effects, we 
may wright the following expressions for the size and angular 
divergence of the photon beam:

\begin{displaymath}
(\sigma_{\mathrm{tot}})^{2}_{x,y} = (\sigma_{\mathrm{e}})^{2}_{x,y} +
\sigma^{2}_{\mathrm{d}} \ , \quad
(\sigma^{\prime}_{\mathrm{tot}})^{2}_{x,y} = 
(\sigma^{\prime}_{\mathrm{e}})^{2}_{x,y} + 
(\sigma^{\prime}_{\mathrm{d}})^{2} \ ,   
\end{displaymath}

\noindent where $\sigma_{\mathrm{d}} =  \sqrt{\lambda 
L_{\mathrm{w}}/(8\pi^{2})}$ is the diffraction limited radiation beam 
size, $\sigma^{\prime}_{\mathrm{d}} = \sqrt{\lambda/(2 
L_{\mathrm{w}})}$ is the diffraction limited radiation beam divergence, 
and $L_{\mathrm{w}}$ is the undulator length \cite{w}. It is frequently 
used in the synchrotron radiation community to express the spectral 
brightness in terms of relative spectral bandwidth of $10^{-3}$.  
The expression for $B$ in terms of photon flux within the central cone 
is \cite{a}:

\begin{displaymath}
B = \frac{(\D N_{\mathrm{cen}}/\D t)(N_{\mathrm{w}}/1000)}
{4\pi^{2}(\sigma_{\mathrm{tot}})_{x}(\sigma_{\mathrm{tot}})_{y}   
(\sigma^{\prime}_{\mathrm{tot}})_{x}
(\sigma^{\prime}_{\mathrm{tot}})_{y}}   
\frac{{\mathrm{photons/s}}}{{\mathrm{mm}}^{2}{\mathrm{mrad}}^{2}
{\mathrm{(0.1 \% BW)}}}  \ .
\end{displaymath}

\subsection{Spatial and spectral filtering of undulator radiation}

Our previous discussion assumed that the synchrotron light striking the 
photosurface was completely coherent in a spatial sense. In such a case 
the number of degrees of freedom $M$ is determined strictly by temporal 
effects. When the wave is not spatially coherent, its spatial structure 
can affect the number of degrees of freedom; at any given time, 
different parts of the photosensitive surface may experience different 
levels of incident intensity. One of the way to overcome this problem 
is based on the idea to use a spatial filter. This  possibility to 
create spatially coherent radiation is important for many experiments 
specifically for our scheme for time-resolved experiments and we will 
discuss in more detail the conditions for the synchrotron light source 
to emit such radiation.

\begin{figure}[tb]
\begin{center}
\epsfig{file=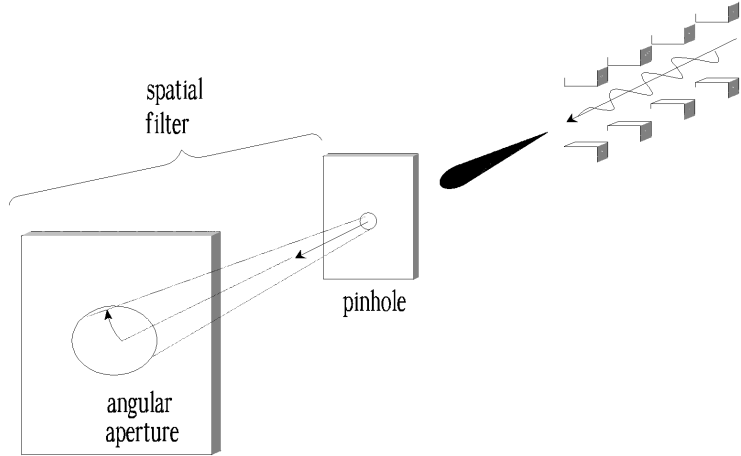,width=0.75\textwidth}
\end{center}
\caption{
Undulator radiation with a pinhole spatial filter} 
\label{fig:pps3} 
\end{figure}

As an example of pinhole spatial filtering, Fig. \ref{fig:pps3} 
illustrates how the technique is used to obtain spatially coherent 
radiation from a periodic undulator \cite{a}.  The limiting condition 
of spatially coherent radiation is a space-angle product $d\cdot\theta 
= \lambda/(2\pi)$, where $d$ is a Gaussian $1/\sqrt{e}$ diameter and 
$\theta$ is the Gaussian half angle \cite{a}.  In general the 
phase space volume of photons in the central radiation cone is larger 
than the limiting condition required for spatial coherence.  Thus, for 
experiments that require spatial coherence, a pinhole and angular 
acceptance aperture are to be introduced.  This pinhole spatial filter 
is used to narrow, or filter, the phase space of transmitted radiation.  
Filtering to $d\cdot\theta = \lambda/(2\pi)$ requires the use of both a 
small pinhole ($d$), and some limitation on $\theta$, such that the 
product is equal to $\lambda/(2\pi)$.

The radiation is transversely coherent when

\begin{displaymath} 
(\sigma_{\mathrm{e}})_{x,y}(\sigma^{\prime}_{\mathrm{e}})_{xy} \ll 
\lambda/(4\pi) \ , \quad \beta \simeq L_{\mathrm{w}} \ .  
\end{displaymath}

\noindent Under these conditions, one has $(\sigma_{x,y})^{2} \ll 
(\sigma_{\mathrm{d}})^{2}$ and $(\sigma^{\prime}_{x,y})^{2} \ll 
(\sigma^{\prime}_{\mathrm{d}})^{2}$ and the spectral-angular dependence 
of the radiation emitted by an electron beam can be approximated by the 
spectral-angular dependence of the radiation emitted by a single  
electron. This limit corresponds to the maximum spectral brightness

\begin{displaymath}
B = \frac{4(\D N_{\mathrm{cen}}/\D t)(N_{\mathrm{w}}/1000)}
{\lambda^{2}}   
\frac{{\mathrm{photons/s}}}{{\mathrm{mm}}^{2}{\mathrm{mrad}}^{2}
{\mathrm{(0.1 \% BW)}}}  \ .
\end{displaymath}

When the electron beam emittance is large,

\begin{displaymath}
(\sigma_{\mathrm{e}})_{x,y}(\sigma^{\prime}_{\mathrm{e}})_{xy} \gg 
\lambda/(4\pi) \ ,  
\end{displaymath}

\noindent the radiation is partially coherent. One can show that the 
influence of the electron beam divergence on the properties of the 
undulator radiation can be neglected when the beta-function is 
large enough. In the region of parameters when 
 
\begin{displaymath}
(\sigma_{x,y}^{\prime})^{2} < (\sigma_{\mathrm{d}}^{\prime})^{2}
\end{displaymath}

\noindent a downstream diaphragm of aperture $d = 
2\sigma_{\mathrm{d}}$ is used for selection of the transversely 
coherent fraction of undulator radiation.  Finally, in this limit, the 
flux of transversely coherent photons into the bandwidth 
$\Delta\omega/\omega = 0.1 \%$ can be estimated simply as

\begin{displaymath}
\frac{\D N_{\mathrm{ph}}}{\D t} \simeq \frac{\lambda^{2} B}{4} \ .  
\end{displaymath}

\noindent These remarks may be useful in estimating the 
conditions under which a spatially coherent plane wavefront may be 
obtained from a noncoherent source.

\begin{figure}[tb]
\begin{center}
\epsfig{file=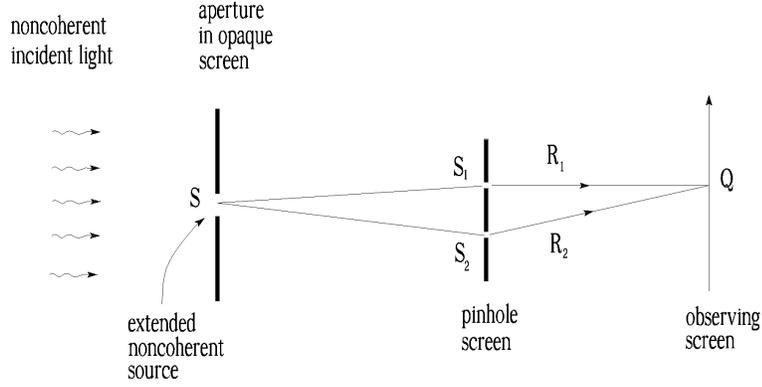,width=0.75\textwidth}
\end{center}
\caption{
Young's interferometer, for discussion of spatial coherence} 
\label{fig:pps10} 
\end{figure}

\subsection{Discussion of spatial coherence}

\begin{figure}[tb]
\begin{center}
\epsfig{file=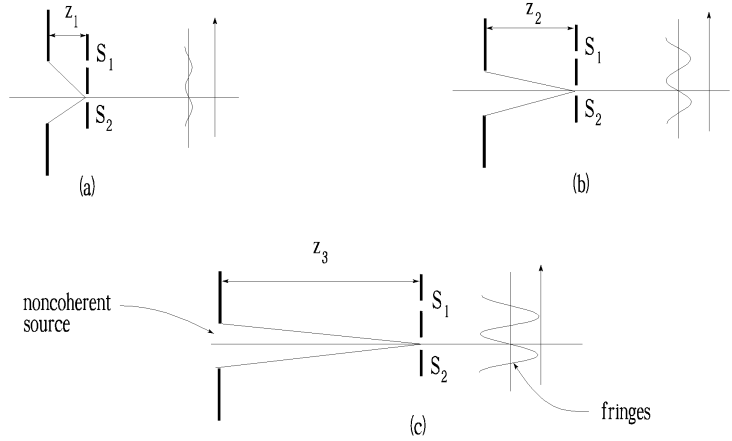,width=0.75\textwidth}
\end{center}
\caption{
Interference-fringe amplitude, in Young's interferometer, for three 
values of noncoherent source-pinhole screen distance} 
\label{fig:pps11} 
\end{figure}

The spatial coherence in a light beam generally has to do with the 
coherence between two points in the field illuminated by light source. 
The meaning of spatial coherence can best be understood with the help 
of Young's two-pinhole experiment (see Fig. \ref{fig:pps10}). 
In its elementary sense, the degree of coherence between the two points 
simply describes the contrast of the interference fringes that are 
obtained when the two points are taken as secondary sources. 
Let a source $S$ illuminate the two pinholes $S_{1}$ and $S_{2}$, as 
shown in Fig. \ref{fig:pps10}. The source is perfectly noncoherent. 
That is to say, no interference fringes can be obtained by placing two 
pinholes in the plane of the source. It was shown, however, that if the 
two slits are placed far enough away from the noncoherent source, 
interference fringes of good contrast can be obtained (see Fig. 
\ref{fig:pps11}).  It is sometimes said that the spatial coherence in 
light beams increases with distance "by mere propagation".  It would be 
nice to find an explanation which is elementary in the sense that we 
can see what is happening physically.    

\begin{figure}[tb]
\begin{center}
\epsfig{file=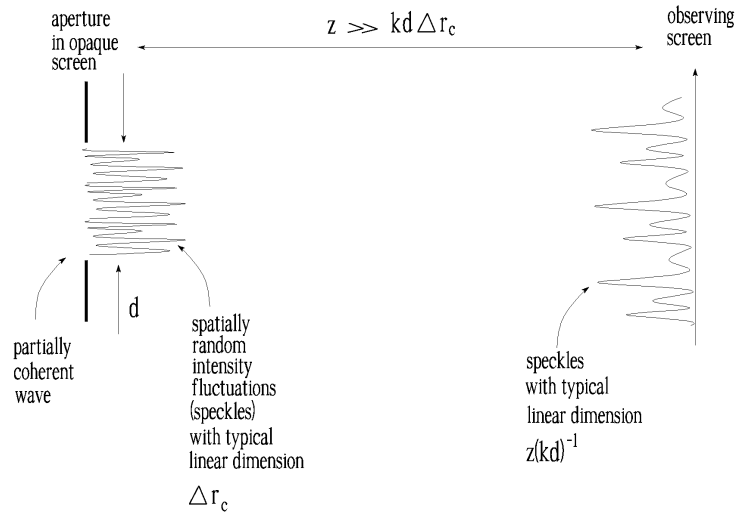,width=0.75\textwidth}
\end{center}
\caption{
Geometry for propagation of spatial coherence} 
\label{fig:pps12} 
\end{figure}

Suppose that a quasimonochromatic wave is incident on an aperture in an 
opaque screen, as illustrated in Fig. \ref{fig:pps12}. In general, this 
wave may be partially coherent. The detailed structure of an optical 
wave undergoes changes at the wave propagates through space. In a 
similar fashion, the detailed structure of the spatial coherence 
undergoes changes, and in this sense the transverse coherence function 
is said to propagate. Knowing the spatial coherence on the aperture, we 
wish to find the spatial coherence on the observing screen at distance 
$z$ beyond the aperture.  Synchrotron radiation is a stochastic object 
and for any synchrotron light beam there exist some characteristic 
linear dimension, $\Delta r_{\mathrm{c}}$, which determines the scale 
of spatially random fluctuations. Fig. \ref{fig:pps12}  illustrates the 
type of spiky speckle pattern on an aperture in an opaque screen.  When 
$\Delta r_{\mathrm{c}} \ll d$, the radiation beyond the aperture is 
partially coherent.  This case is shown in Fig. \ref{fig:pps12}. 
Here $\Delta r_{\mathrm{c}}$ may be estimated as the typical linear 
dimension of speckles.  

\begin{figure}[tb]
\begin{center}
\epsfig{file=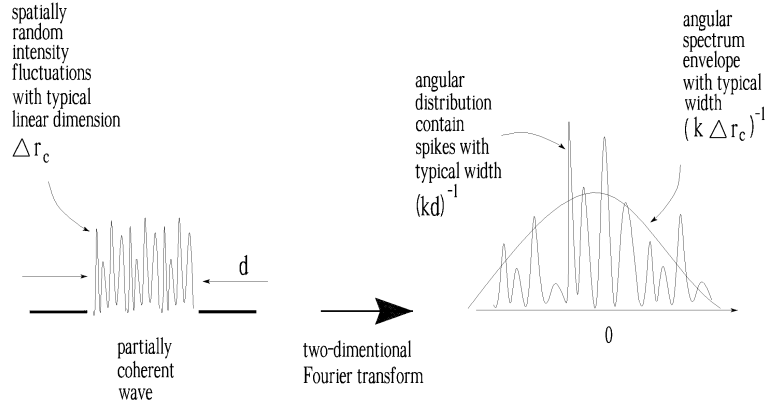,width=0.75\textwidth}
\end{center}
\caption{
Reciprocal width relations of Fourier transform pairs} 
\label{fig:pps14} 
\end{figure}

First we wish to calculate the (instantaneous) intensity distribution 
observed across a parallel plane at distance $z$ beyond the aperture.  
The observed intensity distribution can be found from a two-dimensional 
Fourier transform of the field.  The radiation field across the 
aperture may be presented as a superposition of plane waves, all with 
the same wavenumber $k = \omega_{0}/c$. The value of $k_{\perp}/k$ 
gives the sine of the angle between the $z$ axis and the direction of 
propagation of the plane wave. In the paraxial approximation 
$k_{\perp}/k = \sin \theta \simeq \theta$. If the radiation beyond the 
aperture is partially coherent, a spiky angular spectrum is expected.  
The nature of the spikes in the angular spectrum is easily described in 
Fourier-transform notations. We can expect that the typical width 
of the angular spectrum envelope 
should be of the order of $(k\Delta r_{\mathrm{c}})^{-1}$. Also an 
angular spectrum of the source having transverse size $d$ should 
contain spikes with a typical width of about $(kd)^{-1}$, a consequence 
of the reciprocal width relations of Fourier transform pairs (see Fig. 
\ref{fig:pps14}).  It is the source linear dimension $d$ that 
determines the coherent area of the observed wave $z/(kd)$, but in 
addition the coherence linear dimension $\Delta r_{\mathrm{c}}$ of the 
source influences the distribution of average intensity over the 
observing screen with typical width $z/(k\Delta r_{\mathrm{c}})$.  
Thus, if the screen is placed far enough away from the noncoherent 
source, $z \gg d\Delta r_{\mathrm{c}}/\lambda$, a coherence area of 
large linear dimension can be obtained.     

\subsection{Grating monochromator considered as a spatial filter}

\begin{figure}[tb]
\begin{center}
\epsfig{file=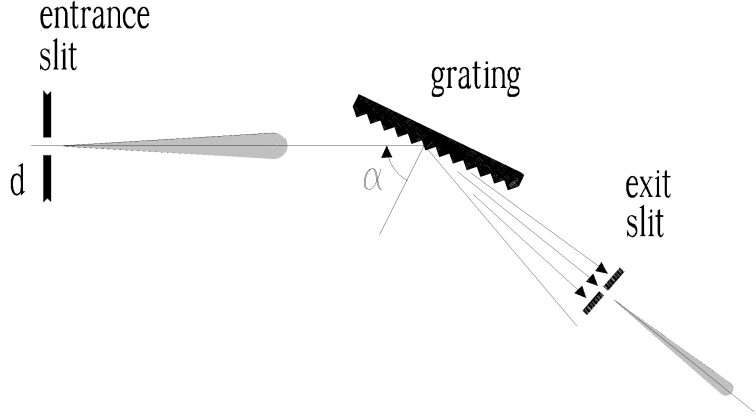,width=0.75\textwidth}
\end{center}
\caption{
Illustration of grating monochromator. The finite size of the entrance 
slit in the dispersive direction sets a limit to the achievable 
resolution } 
\label{fig:pps4} 
\end{figure}

In order to use the radiation from the source it first has to be 
filtered and monochromatized. Finally, to be complete, we should remark 
that in the experimental arrangement shown in Fig. \ref{fig:pps5} there 
is still one other effect which gives spatial filtering.  A 
monochromator is just a spatial filter working in dispersive direction 
only. We would like to show next how it is that a monochromator can act 
like a spatial filter.  Figure~\ref{fig:pps4} shows the grating 
geometry under consideration.  The finite size of the entrance and exit 
slits in the dispersive direction sets a limit to the achievable 
resolution. The slit-width-limited resolution can be obtained directly 
from the grating equation. In particular, the entrance 
slit-width-limited resolution is given by \cite{p}: 

\begin{displaymath}
\Delta\lambda = zd\cos(\alpha)/(nm) \ ,
\end{displaymath}

\noindent where $\alpha$ is the angle of incidence, $N_{\mathrm{g}}$ is 
the number of lines on the grating, $D$ is the width of grating, $n = 
N_{\mathrm{g}}/D$ is the line density, $m$ is the order of the 
diffraction pattern, $d$ is the width of entrance slit, $z$ is the 
entrance slit-grating distance.  We see that the resolution is equal to 

\begin{displaymath}
\frac{\lambda}{\Delta\lambda} = 
\frac{zmN_{\mathrm{g}}\lambda}{Dd\cos(\alpha)} \ . 
\end{displaymath}

\noindent The coefficient $(mN_{\mathrm{g}})^{-1}$ is just the minimal 
uncertainty in the wavelength that can be measured with the given 
grating. The coefficient of proportionality between 
$\lambda/\Delta\lambda$ and $mN_{\mathrm{g}}$ is 
$z\lambda/[Dd\cos(\alpha)]$.  What does all this  mean?  This is an 
example in which transverse coherence plays an important role.  The 
ability to spectrally filter radiation by a monochromator requires 
well-defined phase and amplitude variations of the fields across 
the grating. So in order that we shall have a sharp line in our 
spectrum corresponding to a definite wavelength with an uncertainty 
given by $(mN_{\mathrm{g}})^{-1}$, we have to have a wave transverse 
coherence area of at least the size $D\cos(\alpha)$. If the transverse 
coherence area is too small we are not using the entire grating.  
Suppose we have an entrance slit with aperture $d$ and send a beam of 
radiation at the grating with an aperture $D\cos(\alpha)$. The 
radiation reaching the grating is transversely coherent when the 
following condition for the space-angle product is fulfilled:

\begin{displaymath}
d\cdot \theta = Dd\cos(\alpha)/(2z) \simeq \lambda/(2\pi) \ .
\end{displaymath}

\noindent We expect that only for pure transverse coherence, the 
grating will work right and the resolution $\lambda/\Delta\lambda$ and 
$mN_{\mathrm{g}}$ will be nearly equal. Importantly, in our scheme that 
incorporates an external pinhole spatial filter (see Fig. 
\ref{fig:pps5}), the monochromator does not require an entrance slit.

\subsection{Wave degeneracy parameter of undulator radiation}

Physically the
wave degeneracy parameter describes the average number of photons which 
can interfere, or, according to quantum theory, the number of 
photons in one quantum state (one "mode"). 
We shall deal with two different approaches to the calculation of the 
wave degeneracy parameter. First we consider what can reasonably called 
"fluctuation experiments in the frequency domain" (see subsection 3.6).  
For the frequency domain we use the notion of the wave degeneracy 
parameter $\delta_{\mathrm{W}}$ which is equal to the average number of 
transversely coherent photons radiated by the electron bunch inside the 
spectral interval of coherence $\Delta\omega_{\mathrm{c}}$.  Taking 
into account that the "emittance" of the diffraction limited photon 
beam is equal to

\begin{displaymath}
\min(\epsilon_{x,y}) = \min(2\pi\sigma_{x,y}\sigma^{\prime}_{x,y})
=\frac{\lambda}{2} \ ,  
\end{displaymath}

\noindent we calculate the average number of spatially coherent photons 
radiated within one pulse into the spectral interval of $\Delta\lambda
/\lambda$:

\begin{displaymath}
(\Delta N_{\mathrm{ph}})_{\mathrm{coh}} = \frac{\lambda^{2}}{4}
\frac{\Delta\lambda}{\lambda}\int^{\infty}_{-\infty}B(t)\D t \ , 
\end{displaymath}

\noindent where $B(t)$ is the instantaneous spectral brightness of the 
synchrotron radiation. The value of $\Delta\lambda/\lambda$ is 
connected with the interval of the spectral coherence by the relation:
$\Delta\lambda/\lambda = \Delta\omega_{\mathrm{c}}/\omega_{0}$. Using 
(\ref{eq:e7}), (\ref{eq:e9}) and Parseval's theorem we obtain:

\begin{displaymath}
\frac{\Delta\omega_{\mathrm{c}}}{\omega_{0}} = 
\frac{1}{\omega_{0}}\int\limits^{\infty}_{-\infty}
\mid g_{1}(\Delta\omega)\mid^{2}\D(\Delta\omega) =
\frac{1}{\omega_{0}}\int\limits^{\infty}_{-\infty}
\mid\bar{F}(\Delta\omega)\mid^{2}\D(\Delta\omega) =
\frac{\lambda}{c}\int\limits^{\infty}_{-\infty}
F^{2}(t)\D t \ .
\end{displaymath}

\noindent Using the fact that the instantaneous spectral brightness 
$B(t)$ and the instantaneous value of the beam current $\langle 
I(t)\rangle$ are related by $B(t) \propto \langle I(t)\rangle$, we can 
equivalently write $B(t) = {\mathrm{const}}.\times F(t)$. It is 
convenient to refer to the peak value of spectral brightness 

\begin{displaymath}
\max(B) = {\mathrm{const}}.\times \max(F) = B_{\mathrm{peak}} \ .    
\end{displaymath}
   
\noindent By convention, we represent the time integral of the 
instantaneous spectral brightness in the form 
     
\begin{displaymath}
\int\limits^{\infty}_{-\infty} B(t)\D t = {\mathrm{const}}.\times
\int\limits^{\infty}_{-\infty} F(t)\D t = 
\frac{B_{\mathrm{peak}}}{\max(F)} \ .
\end{displaymath}

\noindent Note that the dependence of the factor

\begin{displaymath}
A = \frac{1}{\max(F)}\int\limits^{\infty}_{-\infty}
F^{2}(t)\D t
\end{displaymath}
  
\noindent on the exact shape of the bunch is rather weak. The results 
are $A = 1$ for the rectangular pulse-shape and $A = 1/\sqrt{2}$ for 
the Gaussian pulse-shape. Thus $A \simeq 1$ is a reasonable 
approximation. Finally, the degeneracy parameter can be estimated 
simply as:

\begin{equation}
\delta_{\mathrm{W}} \simeq \frac{\lambda^{3}B_{\mathrm{peak}}}{4c} \ .
\label{eq:d1}
\end{equation}

What about the other kinds of fluctuation experiments, for example, 
integrated intensity fluctuations?    
For the time domain we use the notion of the wave degeneracy 
parameter $\delta_{\mathrm{W}}$ which is equal to the average number of 
transversely coherent photons radiated by the electron bunch during the 
 coherence time $\tau_{\mathrm{c}}$: 

\begin{displaymath}
(\Delta N_{\mathrm{ph}})_{\mathrm{coh}} = \frac{\lambda^{2}}{4}
\frac{\Delta\lambda}{\lambda}\left(\tau_{\mathrm{c}}
B_{\mathrm{peak}}\right) \ .  
\end{displaymath}

\noindent Finally, the product $\tau_{\mathrm{c}}\Delta\lambda/\lambda$ 
is given by the expression 

\begin{displaymath}
\tau_{\mathrm{c}}\Delta\lambda/\lambda = 
A^{\prime}\lambda/c \ .
\end{displaymath}

\noindent The dependence of the factor $A^{\prime}$ on the exact shape 
of the spectral distribution is rather weak. Let us consider the 
specific case of a Gaussian line profile. The explicit expression for 
the coherence time is $\tau_{\mathrm{c}} = 
\sqrt{\pi}/\sigma_{\mathrm{m}}$. The FWHM bandwidth would be 
$\Delta\lambda/\lambda = 2.36 (\sigma_{\mathrm{m}}/\omega)$ giving, for 
$A^{\prime} = 0.7$. Thus $A^{\prime} \simeq 1$ is 
a reasonable approximation and the wave degeneracy parameter 
$\delta_{\mathrm{W}}$ is given by the expression (\ref{eq:d1}).
The frequency domain calculations 
and time domain calculations give similar results.

Let us present a specific numerical example for the case of a 
third-generation synchrotron light source. The peak spectral 
brightness at a wavelength of 10 nm is equal to

\begin{displaymath}
B_{\mathrm{peak}} \simeq 10^{21}
\frac{{\mathrm{photons/s}}}{{\mathrm{mm}}^{2}{\mathrm{mrad}}^{2}
{\mathrm{(0.1 \% BW)}}}  \ .
\end{displaymath}

\noindent Substituting this number into (\ref{eq:d1}) we obtain that 
the wave degeneracy parameter is about $\delta_{\mathrm{W}} \simeq 
10^{3}$.  Hence in the soft X-ray region of the spectrum we expect 
classically induced fluctuations of photocounts to have a far stronger 
effect than pure quantum shot noise fluctuations. On the other hand, in 
the hard X-ray region of the spectrum ($\lambda \simeq 0.1$ nm), a 
source brightness in excess of  

\begin{displaymath}
B_{\mathrm{peak}} \simeq 10^{24}
\frac{{\mathrm{photons/s}}}{{\mathrm{mm}}^{2}{\mathrm{mrad}}^{2}
{\mathrm{(0.1 \% BW)}}}  
\end{displaymath}

\noindent is required to produce a wave degeneracy parameter 
greater than unity. Since third-generation sources have a peak 
brightness of only this value, we can conclude that in this 
region of the spectrum, the vast majority of third-generation 
synchrotron light sources encountered produce radiation with wave 
degeneracy parameters comparable with unity, and hence noise 
produced by the quantized nature of the radiation is comparable with 
the noise produced by classical fluctuations of the intensity.

A few additional comments are needed in closing this section.
We have concentrated our attention on the wave degeneracy parameter, 
which is a property of the synchrotron radiation falling on the 
photodetector.  That photodetector invariably has a quantum efficiency 
that is less than unity. In addition, the efficiency of the 
monochromator is less than unity, too.  Hence the count degeneracy 
parameter will be even smaller than the wave degeneracy parameter. When 
the integration time is large, $\tau_{0} \gg \tau_{\mathrm{c}}$, 
the parameter $\delta_{\mathrm{c}}$ is given by:

\begin{displaymath}
\delta_{\mathrm{c}} = \eta R_{\mathrm{m}}\delta_{\mathrm{W}} \ .
\end{displaymath}

\noindent where $R_{\mathrm{m}}$ is the monochromator throughput. 
In the opposite case, at $\tau_{0} \ll \tau_{\mathrm{c}}$, the  
 parameter $\delta_{\mathrm{c}}$ is about 

\begin{displaymath}
\delta_{\mathrm{c}} \simeq \eta 
R_{\mathrm{m}}(\Delta\omega_{\mathrm{m}}\tau_{0}) 
\delta_{\mathrm{W}} \ .  
\end{displaymath}

In addition, it is possible that the sample or collecting optics 
may intercept only a fraction of one spatial mode from the source. In 
such a case the count degeneracy parameter may again be smaller than 
the wave degeneracy parameter, as a result of the incomplete capture of 
a spatial mode. Although the minimum value of $M$ is unity, the 
reduction of the energy striking the detector surface must be taken 
into account. In such a case, the count degeneracy parameter must be 
reduced from the normal value by a factor of the ratio of the effective 
measurement area to the coherence area of the incident light.

\section{The signal-to-noise ratio associated with the output of the 
measurement system assumed for the pump-probe experiments}

The principle of operation of the proposed pump-probe scheme is based 
essentially on the classical intensity fluctuations of synchrotron 
radiation.  It is shown above that stochastic fluctuations of the 
classical intensity can influence the statistical properties of the 
photoevents that are observed. Note in particular that the variance 
$\sigma^{2}_{\mathrm{K}}$ of $K$ consists of two distinct terms, each 
of which has a physical interpretation:

\begin{equation} 
\sigma^{2}_{\mathrm{K}} = \frac{\langle K^2\rangle - \langle 
K\rangle^{2}}{\langle K\rangle^{2}} = \frac{1}{\langle K\rangle} + 
\sigma^{2}_{\mathrm{W}} \  
\label{eq:sn1a}
\end{equation}

\noindent The first term is simply the variance of the counts that 
would be observed if the classical intensity were constant and the 
photocounts were purely Poisson. We refer to this contribution to the 
count fluctuations as "quantum noise". The second term, 
$\sigma^{2}_{\mathrm{W}}$, is clearly zero if there are no fluctuations 
of the classical intensity. For instance, in the case of laser light, 
this component would be identically zero, and the count variance would 
be simply that arising from Poisson-distributed counts. When 
synchrotron light is incident on the sample surface, the classical 
fluctuations are nonzero, and the variance of the photocounts is larger 
than that expected for a Poisson distribution by an amount that is 
proportional to the variance of the integrated intensity. This extra 
component of variance of the counts is often referred to as "excess
noise," meaning that it is above and beyond that expected for pure 
Poisson fluctuations \cite{g}.

Our initial goal was to find expressions for the classical variance 
$\sigma^{2}_{\mathrm{W}}$ of the integrated intensity. Also of major 
interest is the signal-to-noise ratio, associated with the variance, 
which provides us with an indication of the magnitude of the 
fluctuations of classical variance relative to the expectation 
value. The general approach to calculating the output signal-to-noise 
ratio of the proposed device will be as follows.  Following 
subtraction of the $\langle K\rangle$ associated with the quantum 
noise (see (\ref{eq:sn1a})), the signal-to-noise ratio takes the 
form\footnote{A complete analysis of the finite averaging should 
include the uncertainties associated with the estimates of all  
average quantities.  For the purpose of simplicity we neglect the 
uncertainty of $\langle K\rangle$.  An assumption that this quantity is 
known much more accurately than $\langle(K - \langle 
K\rangle)^{2}\rangle$ is justified in our case, actually several or 
many different coherence times must be explored, and $\langle K\rangle$ 
is thus observed over many more shots than is $\langle(K - \langle 
K\rangle)^{2}\rangle$ for any one coherence time.}

\begin{equation} 
\frac{S}{N} = 
\frac{\sqrt{N_{\mathrm{shot}}}\sigma^{2}_{\mathrm{W}}\langle 
K\rangle^{2}}{\sqrt{\langle (K - \langle K\rangle)^{4}\rangle - \langle 
(K - \langle K\rangle)^{2}\rangle^{2}}} 
\label{eq:sn1b} 
\end{equation}

\noindent where $\sigma^{2}_{\mathrm{W}}$ is calculated using 
(\ref{eq:e28}), $N_{\mathrm{shot}}$ is the number of independent 
measurements averaged in the accumulator (total number of shots). The 
only requirement for the accuracy of this procedure is that the count 
fluctuations be uncorrelated from shot to shot, a property that does 
hold in our case.

A fully general study of the signal-to-noise ratio would be nontrivial. 
The difficulty arises in simultaneously including the effects on that 
noise of both the classically induced fluctuations and Poisson 
fluctuations of the counts. The full analysis, including both of these 
effects, is a difficult analytical problem. The calculation can be 
performed without great difficulty in two limiting cases, namely, the 
cases of degeneracy parameter very large and very small compared with 
1. When $\delta_{\mathrm{c}} \gg 1$, there are many photoevents 
present in each coherence interval of the wave. The result is a 
"bunching" of the photoevents by the classical intensity fluctuations, 
and an increase of the variance of the counts to the point where the 
classically induced fluctuations are far stronger than the quantum 
noise variations.  When the degeneracy parameter is much larger than 1, 
the expression (\ref{eq:sn1b}) can be replaced by

\begin{equation}
\frac{S}{N} = \frac{\sqrt{N_{\mathrm{shot}}}\langle(K - \langle 
K\rangle)^{2}\rangle}{\sqrt{\langle (K - \langle K\rangle)^{4}\rangle -
\langle (K - \langle K\rangle)^{2}\rangle^{2}}} \quad 
{\mathrm{for}}\quad \delta_{\mathrm{c}} \gg 1 \ . 
\label{eq:sn1c}
\end{equation}

\noindent Our consideration is restricted to the two simplest cases 
from the analytical point of view, namely, the cases of integration 
time $\tau_{0}$ very short and very long compared with the coherence 
time.  For short integration time, the value of the incident 
intensity $I(t)$ is approximately constant over the entire counting 
interval. As a consequence, the integrated intensity $W$ is distributed 
in accordance with the negative exponential distribution 
(\ref{eq:e2a}).  It follows that the $\langle K^{n}\rangle = \langle 
(\alpha W)^{n}\rangle$ can be expressed in terms of the mean as

\begin{displaymath}   
\langle K^{n}\rangle/\langle K\rangle^{n} = n! \ .
\end{displaymath}

\noindent Using the above expression, we find that the signal-to-noise 
ratio is given by the expression: 

\begin{equation}
S/N \simeq  
\sqrt{N_{\mathrm{shot}}/8} \qquad 
{\mathrm{for}} \qquad \delta_{\mathrm{c}} \gg 1, \quad 
\tau_{\mathrm{c}} \gg \tau_{0} \ .  
\label{eq:sn2} 
\end{equation}

\noindent This calculation shows that the fluctuations of the  
$\langle(K - \langle K\rangle)^{2}\rangle$ is about $\sqrt{8} \simeq 3$ 
times larger than the fluctuations of the mean value $\langle 
K\rangle$.

At the opposite extreme, with an integration time much longer than the 
coherence time, the integrated intensity, $W$, is distributed in 
accordance with the Gaussian distribution. It follows that

\begin{equation}
S/N \simeq  
\sqrt{N_{\mathrm{shot}}/2} \qquad 
{\mathrm{for}} \qquad \delta_{\mathrm{c}} \gg 1, \quad 
\tau_{0} \gg \tau_{\mathrm{c}}  \ .  
\label{eq:sn3} 
\end{equation}

We know that for the case when $\delta_{\mathrm{c}} \ll 1$ the 
fluctuations of photocounts are strongly dominated by pure quantum 
noise. Thus the probability of obtaining $K$ counts is, to a good 
approximation, given by the Poisson distribution.  We cannot neglect 
the classically induced fluctuations of the counts when we calculate 
the signal component of the output, but we can neglect them when we 
calculate the noise, simply because their contribution to the noise is 
so small. For the final approximation we note that if the    
integration time $\tau_{0}$ is very short compared with the coherence 
time, the expression for the photocounts degeneracy parameter is equal 
to the average number of detected photoelectrons per pulse 
$\delta_{\mathrm{c}} \simeq \langle K\rangle$. Hence, when 
$\delta_{\mathrm{c}} \ll 1$ and $\tau_{0} \ll \tau_{\mathrm{c}}$ the 
following approximation is valid : $\langle K \rangle \ll 1$. As a 
reminder, recall that the factorial moments of the Poisson distribution 
are given by  

\begin{displaymath}
\langle K(K-1) \cdots (K-n+1)\rangle = \langle K\rangle^{n} \ .
\end{displaymath}

\noindent It follows that

\begin{displaymath}
\langle (K - \langle K\rangle)^{4}\rangle - \langle 
(K - \langle K\rangle)^{2}\rangle^{2}
\simeq \langle K\rangle 
\qquad {\mathrm{for}} \quad \langle K\rangle \ll 1 \ .
\end{displaymath}

\noindent Substituting this result into (\ref{eq:sn1b}) we find                                 
           
\begin{displaymath}
S/N \simeq 
\delta^{3/2}_{\mathrm{c}}\sqrt{N_{\mathrm{shot}}} \qquad
{\mathrm{for}} \qquad \delta_{\mathrm{c}} \ll 1 \ , \quad    
\tau_{0} \ll \tau_{\mathrm{c}} \ .
\end{displaymath}

\noindent Note that this expression is independent of the integration 
time $\tau_{0}$. Therefore, the signal-to-noise ratio is not improved 
by using a longer "characteristic time" of sample dynamics.

\section{Pump-probe technique based on a 
correlation principle}

\begin{figure}[tb]
\begin{center}
\epsfig{file=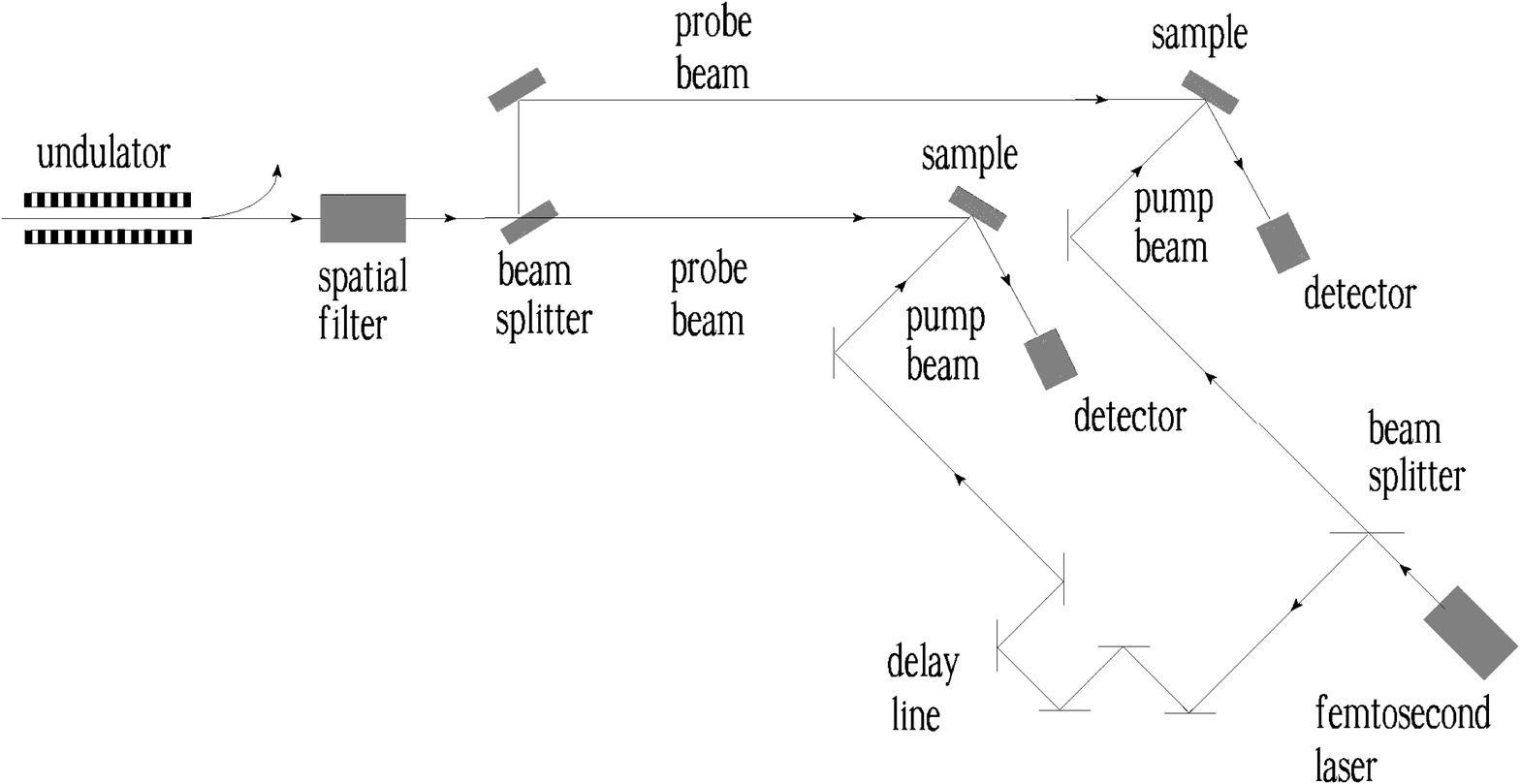,width=0.75\textwidth}
\end{center}
\caption{
Scheme for pump-probe experiments employing correlator of count 
fluctuations} \label{fig:pps17} \end{figure}

\begin{figure}[tb]
\begin{center}
\epsfig{file=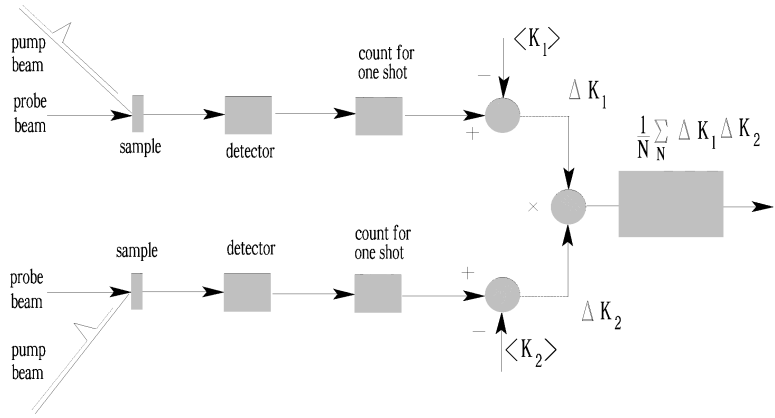,width=0.75\textwidth}
\end{center}
\caption{
Correlation apparatus for analysis of sample dynamics} 
\label{fig:pps16} 
\end{figure}

A more complex instrument can be used to characterize the shape of 
the signal function. The technique we are to describe here  is to use 
two spatially separated samples and record correlation of the count 
fluctuations for each value of the delay $\tau$ between two pump laser 
pulses. The new configuration is sketched in Fig. \ref{fig:pps17}.  In 
this case the transversely coherent probe X-ray beam is splitting and 
incidents on two spatially separated samples.  The detected photoevents 
are than correlated, and the sample dynamics is determined from that 
correlation. Each detector is connected with a separate electronic 
counter. The version of apparatus for measurement of average count 
product depicted in Fig.  \ref{fig:pps16}.  During one radiation pulse, 
each of two detectors registers $K_{1}$ and $K_{2}$ number of 
photocounts, respectively.  After each shot, an electronic scheme 
multiplies these numbers and passes this product to an averaging 
accumulator, where it is added to the previously stored sum of count 
products. Finally, the total sum is divided by the number of shots.  
This result, averaged (shot to shot) count product $\langle 
K_{1}K_{2}\rangle$, contains information about the shape of the signal 
function.  Another advantage of the correlator measurement is the 
possibility to remove the monochromator between spatial filter and 
sample and thus to increase the count degeneracy parameter.

Now our goal is to find the relationship between the 
expected value of the average count product and the sample dynamics. In 
addition, we wish to find the variance of this quantity, so that the 
signal-to-noise ratio associated with the measurement can be determined 
and compared with the similar quantities found in the previous section.

\subsection{The expected value of the count-fluctuation product and its 
relationship to the sample dynamics}
  
By the "count fluctuations" we mean explicitly the difference between 
the actual numbers of counts obtained in one shot at detector 1 and 2 
and the expected values of these two number counts. Thus

\begin{displaymath}
\Delta K_{1} = K_{1} - \langle K_{1}\rangle \ ,
\qquad \Delta K_{2} = K_{2} - \langle K_{2}\rangle \ .
\end{displaymath}

\noindent The averaging accumulator at the output of the system 
depicted in Fig. \ref{fig:pps16} in effect produces an estimate of the 
expected value of the product of the two count fluctuations. Thus we 
are interested in statistical properties of the quantity $\langle 
\Delta K_{1}\Delta K_{2}\rangle$, and particular its mean and variance.

Let us calculate the expected average value of 
$\langle K_{1}K_{2}\rangle$:

\begin{displaymath}
\langle K_{1}K_{2}\rangle
= \sum_{K_{1}=0}^{\infty}\sum_{K_{2}=0}^{\infty}
K_{1}K_{2}P(K_{1},K_{2}) \ ,
\end{displaymath}

\noindent where $P(K_{1},K_{2})$ is the joint probability distribution 
of $K_{1}$ and $K_{2}$. It follows from the basic properties of the 
conditional probabilities that

\begin{displaymath}
P(K_{1},K_{2}) 
= \int\limits^{\infty}_{0}\D W_{1}\int\limits^{\infty}_{0}\D W_{2}
P(K_{1},K_{2}\mid W_{1},W_{2})P(W_{1},W_{2}) \ , 
\end{displaymath}

\noindent where $P(W_{1},W_{2})$ represents the joint probability 
distribution of integrated intensities $W_{1}$ and $W_{2}$. 
The values $W_{1,2}$ can be interpreted  as the portion of X-ray pulse 
energy converted to a detectable form -- photoelectrons for 
example (see section 2.7).  Since $K_{1}$ and $K_{2}$ are  
independent when conditioned by the integrated intensities $W_{1}$ and 
$W_{2}$, respectively, we can write \cite{g}:

\begin{displaymath}
P(K_{1},K_{2}\mid W_{1},W_{2})  
= P(K_{1}\mid W_{1})P(K_{2}\mid W_{2}) \ ,
\end{displaymath}

\noindent Using Mandel's formula (\ref{eq:m}) we can write:

\begin{displaymath}
P(K\mid W) =
\frac{(\alpha W)^{K}}{K!}\exp(-\alpha W) \ . 
\end{displaymath}

\noindent Using these facts, the following expression for the average 
of the count product can now be written

\begin{eqnarray}
& \mbox{} &
\langle K_{1}K_{2}\rangle = \sum^{\infty}_{K_{1}=0}
\sum^{\infty}_{K_{2}=0}K_{1}K_{2}\int\limits^{\infty}_{0}\int\limits^{\infty}_{0}
\frac{(\alpha W_{1})^{K_{1}}}{K_{1}!}\exp(-\alpha W_{1})  
\frac{(\alpha W_{2})^{K_{2}}}{K_{2}!}\exp(-\alpha W_{2})  
\nonumber\\
& \mbox{} &
\times
P(W_{1},W_{2})\D W_{1}\D W_{2} \ .   
\label{eq:k1a}
\end{eqnarray}

\noindent At this point we interchange the orders of summation and 
integration in (\ref{eq:k1a}). Using the relationship 

\begin{displaymath}
\sum^{\infty}_{K=0}K\frac{(\alpha W)^{K}}{K!}\exp(-\alpha W)  
= \alpha W \ ,
\end{displaymath}

\noindent we can express the average of the count product in terms of 
the average of the classical integrated intensities at the two 
detectors \cite{g},

\begin{displaymath}
\langle K_{1}K_{2}\rangle 
= \alpha^{2}\langle W_{1}W_{2}\rangle \ .
\end{displaymath}

This result shows that much can be learnt about the classically induced 
fluctuations by studing the photocount correlation.   
So, the next problem is the description of the correlation of the 
classical integrated intensities  

\begin{displaymath}
W_{1,2} = \int\limits^{\infty}_{-\infty}f_{1,2}(t)I_{1,2}(t)\D t \ ,
\end{displaymath}

\noindent where $f(t)$ is the sample gating function (signal function).
Suppose that the pump pulse interacts with the first and second sample 
at $t = 0$ and $ t = - \tau$, respectively. In this case the gating 
functions are: $f_{1}(t) = f(t), \quad f_{2}(t) = f(t+\tau)$. 
When the average intensities incident on the two samples are constant 
and equal, 
\begin{displaymath} 
\langle I_{1}(t)\rangle = \langle 
I_{2}(t)\rangle = \langle I\rangle = {\mathrm{const.}} \ , 
\end{displaymath}

\noindent the expression for the correlation of count fluctuations 
reduces to the useful form\footnote{This case is particularly 
interesting, because sample dynamics is in the femtosecond 
range, and $\langle I\rangle = {\mathrm{const.}}$ is a good 
assumption.}   

\begin{eqnarray}
& \mbox{} &
\frac{\langle\Delta K_{1}\Delta K_{2}\rangle}{\langle 
K_{1}\rangle\langle K_{2}\rangle} = \frac{\langle\Delta 
W_{1}\Delta W_{2}\rangle}{\langle W_{1}\rangle\langle W_{2}\rangle} 
\nonumber\\
& \mbox{} &
=
\frac{1}{\tau_{\mathrm{ch}}^{2}}\int\limits^{\infty}_{-\infty}\D 
t\int\limits^{\infty}_{-\infty} \D 
t^{\prime}f(t)f(t^{\prime}+\tau)\left[\frac{\langle 
I(t)I(t^{\prime})\rangle}{\langle I\rangle^{2}} - 1\right]  
\nonumber\\
& \mbox{} &
=
\frac{1}{\tau_{\mathrm{ch}}^{2}}\int\limits^{\infty}_{-\infty}\D 
t\int\limits^{\infty}_{-\infty} \D 
t^{\prime}f(t)f(t^{\prime}+\tau)\mid g_{1}(t-t^{\prime})\mid^{2} \ , 
\label{eq:k1}
\end{eqnarray}

\noindent where $\tau_{\mathrm{ch}}$ denotes:

\begin{displaymath}
\tau_{\mathrm{ch}} = \int\limits^{\infty}_{-\infty}f(t)\D t \ .
\end{displaymath}

\noindent Then we remember that the behavior of  
$\mid g_{1}(t-t^{\prime})\mid^{2}$ for  $\tau_{\mathrm{ch}} \gg 
\tau_{\mathrm{c}}$ approaches the behavior of the delta function (here 
$\tau_{\mathrm{ch}}$ is the "characteristic time" of the sample 
dynamics). As a result we obtain

\begin{equation}
\frac{\langle\Delta K_{1}\Delta K_{2}\rangle}{\langle 
K_{1}\rangle\langle K_{2}\rangle} = 
\frac{\tau_{\mathrm{c}}}{\tau_{\mathrm{ch}}^{2}}
\int\limits^{\infty}_{-\infty}
f(t)f(t+\tau)\D t      \qquad
{\mathrm{for}} \qquad 
\tau_{\mathrm{ch}} \gg \tau_{\mathrm{c}} \ .
\label{eq:k2}
\end{equation}

\noindent One can also write (\ref{eq:k2}) as

\begin{displaymath}
\frac{\langle\Delta K_{1}\Delta K_{2}\rangle}{\langle 
K_{1}\rangle\langle K_{2}\rangle} = 
\sigma^{2}_{\mathrm{W}} \frac{
\int\limits^{\infty}_{-\infty}f(t)f(t+\tau)\D t}
{\int\limits^{\infty}_{-\infty}f^{2}(t)\D t} \ ,
\end{displaymath}

\noindent where $\sigma^{2}_{\mathrm{W}}$ is the variance 
of the integrated intensity distribution ( for example see expression 
(\ref{eq:e28})). In the limit of a large photocount 
degeneracy parameter, $\delta_{\mathrm{c}} \gg 1$, we have 
$\sigma^{2}_{\mathrm{W}} \simeq \sigma^{2}_{\mathrm{K}}$.

\begin{figure}[tb]
\begin{center}
\epsfig{file=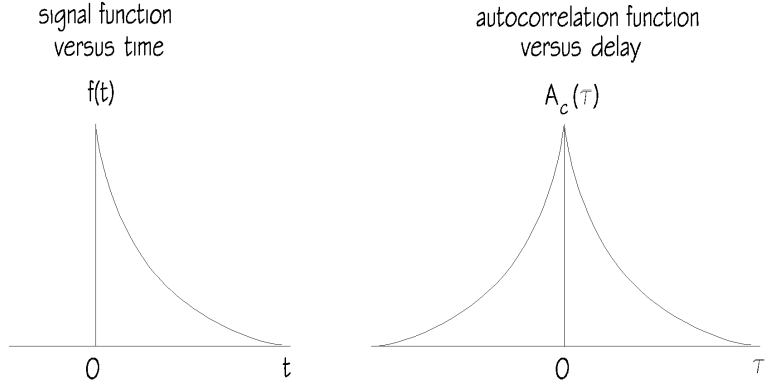,width=0.75\textwidth}
\end{center}
\caption{
Some typical signal versus time and corresponding autocorrelation 
$A_{\mathrm{c}}(\tau)$} 
\label{fig:pps19} 
\end{figure}

The expression 

\begin{displaymath}
A_{\mathrm{c}}(\tau) = \int\limits^{\infty}_{-\infty}f(t)f(t+\tau)\D t
\end{displaymath}

\noindent is called an autocorrelation. 
One immediately recognizes the physical meaning of the autocorrelation 
functions. The Fourier transform of the autocorrelation is 
$\bar{A}_{\mathrm{c}}(\omega)$, related to the Fourier transform of the 
signal by: $\bar{A}_{\mathrm{c}}(\omega) = \mid 
\bar{f}(\omega)\mid^{2}$.  
An autocorrelation is always a symmetric function.
The Fourier transform of the autocorrelation is a real 
function, consistent with a symmetric function in the time domain. 
The question then naturally arises as to exactly what information about 
$f(t)$ can be derived from the measurement of cross-correlation (see 
Fig. \ref{fig:pps19}).  One can see that the proposed correlation 
technique provides the possibility to measure the modulus of the 
Fourier transform of the signal function, while information about its 
phase is missing.  To provide reconstruction of the signal function 
from the autocorrelation function a special iterative technique should 
be used.  Clearly, without loss of generality, we can assume that the 
signal function describing the sample dynamics is zero for $t < 0$ and 
nonzero for $t > 0$.  Additionally we know that at $t = 0$ the signal 
function is maximum.  These relationships help us with the task at 
hand, namely determining the phase of $\bar{f}(\omega)$ from the 
knowledge of its modulus in the majority of cases involving functions 
$f(t)$ of considerable complexity.  Nonetheless, it is possible to find 
cases in which ambiguities exist.

\subsection{The signal-to-noise ratio associated with the output of the 
correlation apparatus}

The signal-to-noise ratio for the proposed device (see Fig. 
\ref{fig:pps16}) can be written in the form:

\begin{equation}
\frac{S}{N} = \frac{\sqrt{N_{\mathrm{shot}}}\langle\Delta 
K_{1}\Delta K_{2}\rangle}{\sqrt{\langle(\Delta 
K_{1}\Delta K_{2})^{2}\rangle - \langle\Delta 
K_{1}\Delta K_{2}\rangle^{2}}}  \ ,  
\label{eq:k3} 
\end{equation}

\noindent where $N_{\mathrm{shot}}$ is the number of independent 
measurements averaged in the accumulator. Let us consider the case when 
the degeneracy parameter is much larger than unity, 
$\delta_{\mathrm{c}} \gg 1$. The fluctuations of the photocounts are 
defined mainly by the classical noise in this case and (\ref{eq:k3}) 
can be reduced to

\begin{equation}
\frac{S}{N} = \frac{\sqrt{N_{\mathrm{shot}}}\langle\Delta 
W_{1}\Delta W_{2}\rangle}{\sqrt{\langle(\Delta 
W_{1}\Delta W_{2})^{2}\rangle - \langle\Delta 
W_{1}\Delta W_{2}\rangle^{2}}} \ ,  
\label{eq:k4} 
\end{equation}

\noindent Remembering that the integrated intensity is distributed in 
accordance with the Gaussian distribution one can derive 
that

\begin{eqnarray}
& \mbox{} &
\frac{\langle W_{1,2}^{2}\rangle}{\langle W_{1}\rangle^{2}} 
= 1 + \sigma^{2}_{\mathrm{W}} \ ,
\nonumber\\
& \mbox{} &
\frac{\langle\Delta W_{1}\Delta W_{2}\rangle}
{\langle W_{1}\rangle\langle 
W_{2}\rangle} = \sigma^{2}_{\mathrm{W}}A_{\mathrm{n}}(\tau) \ , 
\nonumber\\
& \mbox{} &
\frac{\langle W_{1}^{2}W_{2}\rangle}
{\langle W_{1}\rangle^{2}\langle W_{2}\rangle} =
1 + \sigma^{2}_{\mathrm{W}} + 
2\sigma^{2}_{\mathrm{W}}A_{\mathrm{n}}(\tau) \ , 
\nonumber\\
& \mbox{} &
\frac{\langle W_{1}^{2}W_{2}^{2}\rangle}
{\langle W_{1}\rangle^{2}\langle W_{2}\rangle^{2}} =
1 + 2\sigma^{2}_{\mathrm{W}} + 
4\sigma^{2}_{\mathrm{W}}A_{\mathrm{n}}(\tau) 
+ 2\sigma^{4}_{\mathrm{W}}A^{2}_{\mathrm{n}}(\tau) + 
\sigma^{4}_{\mathrm{W}} \ , 
\label{eq:k5a}
\end{eqnarray}
        
\noindent where $A_{\mathrm{n}}(\tau)$ denote 

\begin{displaymath}
A_{\mathrm{n}}(\tau) = 
\frac{\int\limits^{\infty}_{-\infty}f(t)f(t+\tau)\D t} 
{\int\limits^{\infty}_{-\infty}f^{2}(t)\D t} \ .  
\end{displaymath}

\noindent Combining the above expressions, we find that the 
signal-to-noise ratio is given by the expression: 

\begin{equation}
\frac{S}{N} = 
\frac{\sqrt{N_{\mathrm{shot}}}A_{\mathrm{n}}(\tau)}{\sqrt{1 + 
A^{2}_{\mathrm{n}}(\tau)}} \qquad {\mathrm{for}} \qquad 
\delta_{\mathrm{c}} \gg 1 \ .  
\label{eq:k5} 
\end{equation}

Let us analyse the signal-to-noise ratio taking into 
account the quantum effects. When calculating the signal component 
(numerator of (\ref{eq:k3})), we can take into account classically 
induced fluctuations of the counts only. This can be done due to the 
fact that the quantum fluctuations of the counts at the outputs of two 
detectors are statistically independent. General calculations of the 
noise fluctuations associated with the output of the correlation 
apparatus (denominator of (\ref{eq:k3})) should include both the 
classical and the quantum effects. Taking into account these 
considerations we can derive the following relations:

\begin{eqnarray}
& \mbox{} &
\frac{\langle K_{1}^{2}K_{2}^{2}\rangle}{\langle 
K_{1}\rangle^{2}\langle K_{2}\rangle^{2}} =
\frac{\langle(\alpha^{2}W_{1}^{2} + \alpha W_{1})(\alpha^{2}W_{2}^{2} + 
\alpha W_{2})\rangle}{\langle\alpha W_{1}\rangle^{2}
\langle\alpha W_{2}\rangle^{2}} \ ,    
\nonumber\\
& \mbox{} &
\frac{\langle K_{1}^{2}K_{2}\rangle}{\langle 
K_{1}\rangle^{2}\langle K_{2}\rangle} =
\frac{\langle(\alpha^{2}W_{1}^{2} + \alpha W_{1}) 
\alpha W_{2}\rangle}{\langle\alpha W_{1}\rangle^{2}
\langle\alpha W_{2}\rangle} \ ,    
\nonumber\\
& \mbox{} &
\frac{\langle K_{1,2}^{2}\rangle}{\langle 
K_{1,2}\rangle^{2}} =
\frac{\langle\alpha^{2}W_{1,2}^{2} + \alpha W_{1,2}\rangle}{\langle
\alpha W_{1,2}\rangle^{2}} \ .    
\label{eq:k5b}
\end{eqnarray}

\noindent We illustrate the technique for calculations of the above 
expressions for the value of $\langle K_{1}^{2}K_{2}^{2}\rangle$:

\begin{eqnarray}
& \mbox{} &
\langle K_{1}^{2}K_{2}^{2}\rangle =
\sum_{K_{1}=0}^{\infty}\sum_{K_{2}=0}^{\infty}
K_{1}^{2}K_{2}^{2}P(K_{1},K_{2})
=
\nonumber\\
& \mbox{} &
\int\limits_{0}^{\infty}\D W_{1}\int\limits \D W_{2}P(W_{1},W_{2})
\times
\nonumber\\
& \mbox{} &
\left[\sum_{K_{1}=0}^{\infty}
K_{1}^{2}\frac{(\alpha W_{1})^{K_{1}}}{K_{1}!}\exp(-\alpha W_{1})  
\right]\times\left[\sum_{K_{2}=0}^{\infty}
K_{2}^{2}\frac{(\alpha W_{2})^{K_{2}}}{K_{2}!}\exp(-\alpha W_{2})  
\right] =
\nonumber\\
& \mbox{} &
\int\limits_{0}^{\infty}\D W_{1}\int\limits \D W_{2}P(W_{1},W_{2})
(\alpha^{2}W_{1}^{2} + \alpha W_{1}) 
(\alpha^{2}W_{2}^{2} + \alpha W_{2}) =
\nonumber\\
& \mbox{} &
\langle(\alpha^{2}W_{1}^{2} + \alpha W_{1}) 
(\alpha^{2}W_{2}^{2} + \alpha W_{2})\rangle \ . 
\label{eq:k5c}
\end{eqnarray}

\noindent We will consider only the special case when the degeneracy 
parameter is much less than 1. Using the approximation 
$\delta_{\mathrm{c}} \ll 1$ we obtain $(\delta_{\mathrm{c}})_{1,2} 
\simeq \langle K_{1,2}\rangle$.  Under such condition one  would have, 
from (\ref{eq:k5b}), the  variance  

\begin{displaymath}
\langle (\Delta K_{1}\Delta K_{2})^{2}\rangle - 
\langle\Delta K_{1}\Delta K_{2}\rangle^{2}  
\simeq
\langle K_{1}K_{2}\rangle \ , 
\end{displaymath}

\noindent and the signal-to-noise ratio would be

\begin{equation}
\frac{S}{N} = 
\sqrt{N_{\mathrm{shot}}}\delta_{\mathrm{c}}
A_{\mathrm{n}}(\tau) \qquad 
{\mathrm{for}} \qquad \delta_{\mathrm{c}} \ll 1 \ .  
\label{eq:k6} 
\end{equation}
 
\noindent The analysis of the obtained asymptotic expressions 
(\ref{eq:k5}) and (\ref{eq:k6}) for the signal-to-noise ratio allows 
one to make the following conclusions:

\noindent (i) The signal-to-noise ratio depends on the degeneracy 
parameter only when $\delta_{\mathrm{c}}$ is much less that 1. In this 
case the number of independent measurements for a given signal-to-noise 
ratio is proportional to $(1/\delta_{\mathrm{c}})^{2}$.

\noindent (ii) The signal-to-noise ratio is proportional to 
$A_{\mathrm{n}}(\tau)$. To hold the signal-to-noise ratio constant, the 
number of independent measurements must be proportional to 
$A^{-2}_{\mathrm{n}}(\tau)$.

\subsection{Other autocorrelation techniques}

\begin{figure}[tb]
\begin{center}
\epsfig{file=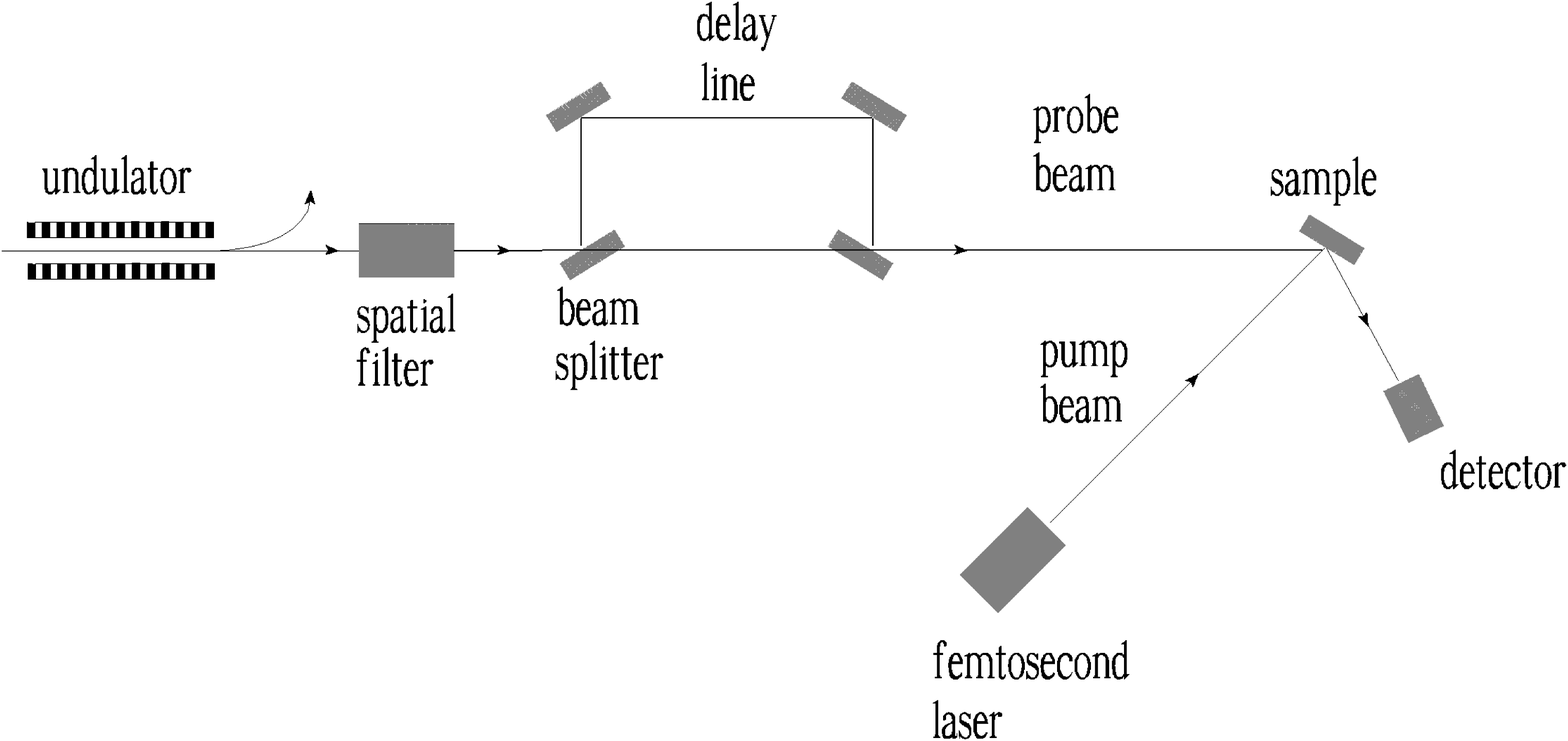,width=0.75\textwidth}
\end{center}
\caption{
Scheme for autocorrelation pump-probe experiments employing two probe 
beams originating from the splitting of a single X-ray beam produced by 
an undulator (with one partial beam delayed relative to the other)} 
\label{fig:pps18} 
\end{figure}

There are other techniques to provide the possibility to 
measure the integrated intensity autocorrelation. They differ from each 
other in the beam splitters, the delay line and the detection 
system. We will now describe two new approaches. An alternative 
technique to generate $A_{\mathrm{n}}(\tau)$ is to use one pump beam 
and two probe beams. In the setup shown in Fig. \ref{fig:pps18} the 
pump pulse perturbs the sample at time $ t = 0$. A long X-ray pulse 
passes a beam splitter, generating two pulses travelling different 
paths. After the two paths are recombined, the two pulses impinge on 
the sample at a time difference $\tau$ which can be varied. The 
photocounts are recorded by a detector and the variance 
$\sigma^{2}_{\mathrm{K}}$ is studied as a function of $\tau$. 

Now we describe the basic principles of autocorrelation measurement 
with this arrangement. In the limit of a large photocount 
degeneracy parameter the expression for the variance of counts reduces 
to the form ($\langle I_{1}\rangle = \langle I_{2}\rangle$ =  const.):

\begin{eqnarray}
& \mbox{} &
\frac{\langle(\Delta K)^{2}\rangle}{\langle 
K\rangle^{2}} = \frac{\langle(\Delta 
W)^{2}\rangle}{\langle W\rangle^{2}} 
\nonumber\\
& \mbox{} &
=
\frac{1}{\tau_{\mathrm{ch}}^{2}}\int\limits^{\infty}_{-\infty}\D 
t\int\limits^{\infty}_{-\infty} \D 
t^{\prime}f(t)f(t^{\prime})\left\{\frac{\langle 
[I_{1}(t)+I_{2}(t+\tau)][I_{1}(t^{\prime})+I_{2}(t^{\prime}+\tau)]
\rangle}{\langle (I_{1} + I_{2})\rangle^{2}} - 1\right] 
\nonumber\\ 
& \mbox{} & 
= 
\frac{1}{2\tau_{\mathrm{ch}}^{2}}\int\limits^{\infty}_{-\infty}\D 
t\int\limits^{\infty}_{-\infty} \D 
t^{\prime}f(t)f(t^{\prime})\left[\mid g_{1}(t-t^{\prime})\mid^{2}  
+ \mid g_{1}(t-t^{\prime}+\tau)\mid^{2}\right] \ .
\label{eq:k7}
\end{eqnarray}

\noindent Remembering that the behavior of $\mid 
g_{1}(t-t^{\prime})\mid^{2}$ for $\tau_{\mathrm{ch}} \gg 
\tau_{\mathrm{c}}$ approaches the behavior of the delta function we 
obtain\footnote{We can assume that  
$\tau \gg \tau_{\mathrm{c}}$, so that no interference is possible 
between waves of different probe beams. In this case the 
contribution to the light intensity is obtained by a suitable summation 
of the intensity contributions. Since we assume that 
$\tau_{\mathrm{ch}} \gg \tau_{\mathrm{c}}$, there is relatively little 
loss of generality by making this additional assumption.}

\begin{equation}
\sigma^{2}_{\mathrm{W}}(\tau) = 
\frac{1}{2}\left[\frac{\tau_{\mathrm{c}}}{\tau_{\mathrm{ch}}^{2}}
\int\limits^{\infty}_{-\infty}
f^{2}(t)\D t\right]\left[
1 + A_{\mathrm{n}}(\tau)\right] \qquad
{\mathrm{for}} \qquad 
\tau_{\mathrm{ch}} \gg \tau_{\mathrm{c}} \ .
\label{eq:k8}
\end{equation}

\noindent One can also write (\ref{eq:k8})

\begin{displaymath}
\sigma^{2}_{\mathrm{W}}(\tau) = 
\frac{1}{2}\sigma^{2}_{\mathrm{W}}(0)
\left[1 + A_{\mathrm{n}}(\tau)\right] \ ,
\end{displaymath}

\noindent where  $\sigma^{2}_{\mathrm{W}}(0)$ is the variance of the 
integrated intensity distribution for the single probe beam:

\begin{displaymath}
\sigma^{2}_{\mathrm{W}}(0) =
\frac{\tau_{\mathrm{c}}}{\tau_{\mathrm{ch}}^{2}}
\int\limits^{\infty}_{-\infty}
f^{2}(t)\D t \ .
\end{displaymath}

\noindent At $ \tau = 0$, the peak value of the function  
$\sigma^{2}_{\mathrm{W}}(\tau)$ is $\sigma^{2}_{\mathrm{W}}(0)$. 
For large delays 
compared to the "characteristic time" of the sample dynamics, the cross 
product term vanishes, leaving a background of 
$\sigma^{2}_{\mathrm{W}}(\infty) = \sigma^{2}_{\mathrm{W}}(0)/2$. 
The variance as a function of delay time has thus a peak to background 
of 2 to 1. The measurement leading to $\sigma^{2}_{\mathrm{W}}(\tau)$ 
we can refer to as the integrated intensities autocorrelation with 
background, as opposed to the background free autocorrelation described 
in the previous subsection (see Fig. \ref{fig:pps17}).

The most important and critical area of R\&D for time-resolved 
experiments utilizing the autocorrelator will be the development of 
X-ray beam splitters and delay lines. Considerable work will need to be 
undertaken to make them routinely used optical elements.

\begin{figure}[tb]
\begin{center}
\epsfig{file=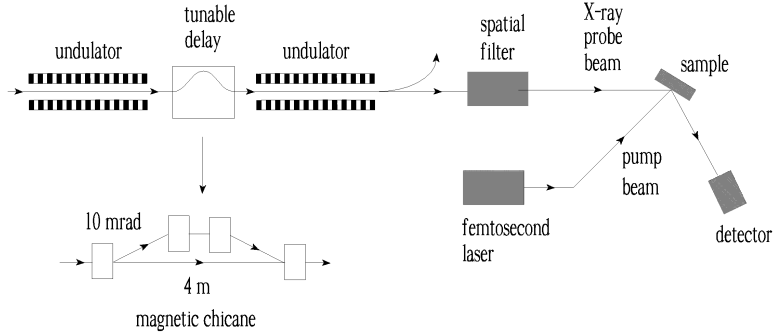,width=0.75\textwidth}
\end{center}
\caption{
Scheme for autocorrelation pump-probe experiments employing two probe 
beams that are generated by the same electron bunch but from two 
undulators} 
\label{fig:pps15} 
\end{figure}

The technique to use two probe X-ray beams that are generated by the 
same electron bunch but from two insertion devices (again with one beam 
delayed with respect to the other) may be  a more promising approach.
The concept is very interesting in view of the critical synchronization 
of probe beams (see Fig. \ref{fig:pps15}). Using this technique 
precisely controlled temporal resolution down to several femtoseconds 
should be achievable.  To delay the electron beam, a small magnetic 
chicane has to be applied.  The trajectory of the electron beam in the 
chicane has the shape of an isosceles triangle with the base equal to 
$L$. The angle adjacent to the base, $\theta$, is considered to be 
small. The optimal parameters of the chicane can be calculated in the 
following way. The electron beam delay line needed for time resolved 
experiments has to satisfy two requirements.  First, the electron beam 
extra path length, $L\theta^{2}/2$, must be long enough to reach a 
delay time of about 100 femtoseconds. A first question that arises is: 
why do we want to use 100 fs delay time? Here we direct the attention 
of the reader to the fact that we assume to use 5-10 fs visible pump 
pulses. On the other hand we assume that the pump pulse can be 
approximated by a delta function. One can never hope to achieve a 
visible laser pulse duration ten times shorter than the present. This 
limitation prevents the application of this technique to the 
timescale shorter than 100 fs.  

Second, the 
effect of suppressing the beam density modulation due to the presence 
of the energy spread in the electron beam should be avoided in order to 
preserve the correlation of the radiation field within the delay time. 
The energy spread suppresses significantly the microbunching when 
$\omega_{0}L\theta^{2}\Delta\gamma/(c\gamma)$
is larger than or equal to unity. Parameters in our case are:  
$\Delta\gamma/\gamma \simeq 10^{-3}$, $L\theta^{2} \simeq 60 
\mu{\mathrm{m}}$. The delay time and energy spread in the electron beam 
dictates the choice of synchrotron radiation wavelengths. These 
results demonstrate that this scheme is adequate for the VUV spectral 
range, but cannot be used to produce X-ray probe beams. The situation 
is quite different for the first scheme of autocorrelator (see Fig. 
\ref{fig:pps18}).  Since in this case the probe beams are produced by 
X-ray splitter and delay line, the energy spread limitation does not 
exist at all.   

\section{Concluding remarks}

A general objective in the development of synchrotron radiation sources 
is to produce radiation that is brighter than that from 
existing sources, or to produce radiation that comes in shorter pulses. 
Current machines produce bunch lengths in the 100-ps regime. Many 
experiments exploit this time dependence.
Light-triggered time-resolved 
studies with 100 ps resolution are now frequently performed at 
synchrotron light sources.  
However, 100 picoseconds is longer than the time associated with many 
interesting physical phenomena. The storage ring technology itself 
approaches its limits of performance with respect to minimal pulse 
duration.  No practical means have been found to allow bunches of less 
than  a ten picoseconds duration to be stored. 
A proposed in this paper 
schemes for fast time-resolved experiments could offer the possibility 
to overcome this limitation. It was shown that the 
femtosecond timescale associated with intensity fluctuations in 
synchrotron light pulses make them well suited for time-resolved 
studies. The new concept eliminates the need 
for ultrashort X-ray pulses.
This development in synchrotron based pump-probe experiments 
allows us to investigate phenomena in the time range down 
to 100 fs. It is the author's hope that the general principles of 
pump-probe techniques based on statistical properties of synchrotron 
light, presented in this paper, may help in stimulating further inquiry 
and result in success in many exciting applications, some of which can 
now only be barely envisioned.

\section*{Acknowledgments}

We thank  J. Feldhaus, T. Moeller and D. Novikov for many useful 
discussions.  We thank J.R.  Schneider and D.  Trines for their 
interest in this work.

\end{document}